\documentclass[11pt,a4paper]{article}
\usepackage[dvipdfmx]{graphicx}
\usepackage[dvipdfmx]{color} 
\usepackage{amsmath,amssymb,amsfonts,color,scalefnt}
\usepackage{epsfig,amsthm}
\usepackage{soul}
\usepackage[makeroom]{cancel}
\usepackage{mathrsfs}
\usepackage{epstopdf}
\usepackage[utf8]{inputenc}
\usepackage{slashed}
\usepackage{graphicx}
\usepackage{pict2e}
\usepackage{tikz}
\usepackage{cite}
\usepackage{float}
\usepackage[percent]{overpic}
\usepackage{multirow}
\usepackage{physics}
\usepackage{booktabs}
\usepackage{mwe}
\usepackage [autostyle, english = american]{csquotes}
\usepackage{bbm}
\usepackage{todonotes}
\MakeOuterQuote{"}

\newcommand{\beq}{\begin{eqnarray}}
\newcommand{\eeq}{\end{eqnarray}}
\newcommand{\ba}{\begin{eqnarray}}
\newcommand{\ea}{\end{eqnarray}}
\newcommand{\be}{\begin{equation}}
\newcommand{\ee}{\end{equation}}
\newcommand{\bpmatrix}{\begin{pmatrix}}
\newcommand{\epmatrix}{\end{pmatrix}}

\newcommand{\comment}[1]{\ignorespaces}

\begin{document}

\title{
	\vspace*{-3cm}
	\phantom{h} \hfill\mbox{\small KA-TP-01-2022}
	\vspace*{0.7cm}
\\[-1.1cm]
	\vspace{15mm}   
	\textbf{One-loop Corrections to the Higgs Boson Invisible
          Decay in a Complex Singlet Extension of the SM \\[4mm]}} 
\date{}
\author{
Felix Egle$^{1\,}$\footnote{E-mail: \texttt{felix.egle@kit.edu}},
Margarete M\"{u}hlleitner$^{1\,}$\footnote{E-mail:
	\texttt{margarete.muehlleitner@kit.edu}},
Rui Santos$^{2,3\,}$\footnote{E-mail:  \texttt{rasantos@fc.ul.pt}}
Jo\~ao Viana$^{2\,}$\footnote{E-mail:  \texttt{jfvvchico@hotmail.com}},
\\[9mm]
{\small\it
$^1$Institute for Theoretical Physics, Karlsruhe Institute of Technology,} \\
{\small\it Wolfgang-Gaede-Str. 1, 76131 Karlsruhe, Germany.}\\[3mm]
{\small\it
$^2$Centro de F\'{\i}sica Te\'{o}rica e Computacional,
    Faculdade de Ci\^{e}ncias,} \\
{\small \it    Universidade de Lisboa, Campo Grande, Edif\'{\i}cio C8
  1749-016 Lisboa, Portugal} \\[3mm]
{\small\it
$^3$ISEL -
 Instituto Superior de Engenharia de Lisboa,} \\
{\small \it   Instituto Polit\'ecnico de Lisboa
 1959-007 Lisboa, Portugal} \\[3mm]
}
\maketitle

%
%

\begin{abstract}
The search for dark matter (DM) at colliders is founded on the idea of looking for something invisible. There are searches 
based on production and decay processes where DM may reveal itself as missing energy. If nothing is found, our best
tool to constrain the parameter space of many extensions of the
Standard Model (SM) with a DM candidate is the Higgs boson. As the 
measurements of the Higgs couplings become increasingly precise, higher-order corrections will start to play a major role.
The tree-level contribution to the invisible decay width provides information about the portal coupling. Higher-order corrections
also gives us access to other parameters from the dark sector of the Higgs potential that are not present in the tree-level amplitude.
In this work we will focus on the complex singlet extension of the SM in the phase with a DM candidate. We calculate the 
one-loop electroweak corrections to the decay of the Higgs boson into
two DM particles. We find that the corrections are stable and of the
order of a few percent.  The present measurement of the Higgs
invisible branching ratio, BR$(H \to$ invisible $) < 0.11$, already
constrains the parameter space of the model at leading order. We
expect that by the end of the LHC the experimental measurement will
require the inclusion of the
electroweak corrections to the decay in order to match
the experimental accuracy. Furthermore, the only competing process, which is direct
detection, is shown to have a cross section below the 
neutrino floor.
\end{abstract}

\newpage
\maketitle

\thispagestyle{empty}
\vfill
\newpage

\section{Introduction}
\label{sec:intro}

The search for dark matter (DM) has replaced the search for the Higgs boson as the main goal of particle physicists. In fact, since the Higgs
has been discovered at the Large Hadron Collider (LHC) by the ATLAS~\cite{Aad:2012tfa} and CMS~\cite{Chatrchyan:2012ufa} collaborations, 
and the Higgs couplings have been measured with great precision, the attention has turned to the outstanding problems of the Standard Model (SM).
The search for DM is certainly on the top of the list especially
because at this point we cannot even be sure if 
it comes in the form of
an elementary particle. Therefore, even if collider physics is not the place to prove a DM candidate exits, it can help us by hinting at
some particular directions even if only by excluding the parameter space of particular models. The Higgs invisible decay measurements
are probably one of best quantities to probe the dark sector of particular models. The branching ratio of Higgs to invisible is now bounded 
to below 11\% by ATLAS~\cite{ATLAS:2019cid}. This number will improve both in the next LHC run and in the high luminosity stage. This
increasing precision will take us further inside the dark sector of the models. 

In this work we discuss the Higgs invisible decay in the Complex
Singlet extension of the SM (CxSM) which amounts to the addition 
of a complex scalar singlet to the known SM fields while keeping the
SM gauge symmetries. While the tree-level decay of the Higgs 
into DM involves only the portal coupling, the one-loop corrections to
the decay give us access to the quartic coupling of the singlet field. 
Therefore, the one-loop result gives us a more complete understanding
of the Higgs potential. There is a competing/complementary
measurement which is the one given by the direct detection
process. 
The DM-nucleon cross section is only relevant at one-loop due to
a cancellation that renders the tree-level cross section proportional to the DM velocity and therefore negligible~\cite{Gross:2017dan,Azevedo:2018oxv}. 
The one-loop corrections to the direct detection process were calculated in~\cite{Azevedo:2018exj, Glaus:2020ihj} and compared to the latest 
experimental results from XENON~\cite{XENON:2018voc}. We will discuss
the interplay between direct detection and the branching ratio 
of the invisible Higgs decay including the electroweak
corrections in both processes.

Our analysis will be performed taking into account the most relevant theoretical and experimental constraints on the model. These are collider constraints
and also DM constraints. We will then calculate the next-to-leading order (NLO) electroweak corrections to the invisible decay width of the SM-like Higgs boson
using several renormalization schemes. Once the allowed parameter
space is found, the NLO result will be compared with the leading order
(LO) one. The final goal is to understand if the NLO Higgs branching
ratio into two DM particles can be larger than  the experimentally
measured value for some regions of the parameter space. Moreover, as the new data
will become available both at the next LHC run and at the high luminosity
stage the Higgs coupling measurements will be more precise and the
theoretical calculations need to match this precision.

The outline of the paper is as follows. In section \ref{sec:model}, we will introduce the CxSM together with our notation. Section~\ref{ch:renormalization} 
is dedicated to the description of the different renormalization schemes used in this work. Section~\ref{sec:constraints} discusses the experimental
and theoretical constraints on the model. In section \ref{sec:resdis}, the results are presented and discussed. Our conclusions are collected in
section~\ref{sec:conc}. 
Finally, there are two appendices, the first one where the results for
the scalar pinched self-energies are presented and the second one
where we discuss the minima of the CxSM potential.

\section{The CxSM Potential}
\label{sec:model}

In this section we introduce the version of the CxSM used in this
work. The model is a simple extension of the SM by the addition of a
complex singlet field with zero isospin and zero hypercharge.  
As a singlet for the SM gauge group, the scalar field appears only in
the Higgs potential. The SM Higgs couplings will be, however, modified
by the rotation angle from the matrix that relates 
the scalar gauge eigenstates with their mass eigenstates. The doublet
field $\Phi$ and the singlet field $\mathbb{S}$ are defined as
\begin{align}
\label{eq:scalar_vev_structure}
 \Phi= \begin{pmatrix}
G^+\\\frac{1}{\sqrt{2}} \left( v+H+iG^0 \right)
\end{pmatrix}, \:
\mathbb{S}=\frac{1}{\sqrt{2}}(v_{S}+S+i(v_{A}+A)),
\end{align}
where $H$, $S$ and $A$ are real scalar fields and $G^+$ and $G^0$ are
the Goldstone bosons for the $Z$ and $W^{\pm}$ bosons. 
The $v$, $v_A$ and $v_S$ are the vacuum expectation values (VEVs) of
the corresponding fields and can all be, in general, non-zero in which
case mixing between all three scalar fields arises. 
We will, however, focus on a model where a DM candidate is generated by forcing the potential to be invariant under a symmetry,  unbroken by the vacuum. We choose
to impose invariance of the potential under two separate $\mathbb{Z}_2$ symmetries acting on $S$ and $A$, that is, $S \rightarrow -S$ and $A \rightarrow -A$. The resulting renormalizable potential is 
\begin{align}\label{eq:scalar_potential}
\begin{split}
V=\frac{m^2}{2} \Phi^{\dagger}  \Phi + \frac{\lambda}{4}\left(  \Phi^{\dagger}  \Phi \right)^2 + \frac{\delta_{2}}{2} \Phi^{\dagger}  \Phi|\mathbb{S}|^2  +\frac{b_{2}}{2}|\mathbb{S}|^2+\frac{d_2}{4}|\mathbb{S}|^4 + \left( \frac{b_1}{4}\mathbb{S}^2+c.c.\right),
\end{split}
\end{align}
where all constants are real. By choosing $v_A=0$, the $A \rightarrow -A$ symmetry remains unbroken and $A$ is stable, becoming the DM candidate of the model. The other  $\mathbb{Z}_2$ symmetry is broken 
since $v_S \neq 0$ which leads to mixing between $S$ and $H$. The mass eigenstates of the CP-even field $h_i$ ($i=1,2$) relate to the gauge eigenstates $H$ and $S$ through 
\begin{align}
\label{eq:rotationdef}
\begin{pmatrix}
h_1 \\ h_2
\end{pmatrix}
=R_{\alpha}
\begin{pmatrix}
H \\ S \\
\end{pmatrix},
\end{align}
where the rotation matrix is given by 
\begin{align}
\label{eq:DefRotationmatrix}
R_\alpha=\begin{pmatrix}
\cos \alpha & \sin\alpha  \\
-\sin \alpha & \cos \alpha  \\
\end{pmatrix} \equiv
\begin{pmatrix}
c_{\alpha} & s_{\alpha} \\
-s_{\alpha} & c_{\alpha} \\
\end{pmatrix} .
\end{align}
The mass matrix in the gauge basis $(H,S)$ is given by
\begin{align}
\label{eq:massmatrix}
\mathcal{M}= \begin{pmatrix}
\frac{v^2\lambda}{2} & \frac{\delta_{2}vv_{S}}{2} \\
\frac{\delta_{2}vv_{S}}{2} & \frac{d_{2}v_{S}^2}{2} \\
\end{pmatrix} + \begin{pmatrix}
\frac{T_{1}}{v} & 0 \\
0 & \frac{T_{2}}{v_{S}}\\
\end{pmatrix},
\end{align}
where the tadpole parameters $T_1$ and $T_2$ are defined via the
minimisation conditions, 
\begin{subequations}
\label{eq:minimizationcond}
\begin{align}
\frac{\partial V}{\partial v}  \equiv T_1 \; \Rightarrow \; \frac{T_{1}}{v}&= \frac{m^2}{2}+\frac{\delta_2v_{S}^2}{4}+ \frac{v^2\lambda}{4},\\
\frac{\partial V}{\partial v_S} \equiv T_2 \; \Rightarrow \; \frac{T_{2}}{v_{S}}&= \frac{b_1+b_2}{2}+\frac{\delta_2v^2}{4}+ \frac{v_{S}^2d_2}{4} \, ,
\end{align}
\end{subequations}
and at tree level, the minimum conditions are $T_i = 0$ ($i=1,2$). The mass of the DM candidate $A$ is given by
\begin{align}
m_{A}^2=\frac{-b_1+b_2}{2}+\frac{\delta_2v^2}{4}+ \frac{v_{S}^2d_2}{4}=-b_1 + \frac{T_2}{v_S},
\end{align}
while the remaining mass eigenstates are obtained via 
\begin{align}\label{eq:massmatrixdiag}
D_{hh}^{2} \equiv R_{\alpha} \mathcal{M} R^{T}_{\alpha} \;,\qquad
  D_{hh}^2=\mbox{diag}(m_{h_1}^2,m_{h_2}^2) \;.
\end{align}
Therefore, the scalar spectrum of the CxSM consists of two Higgs
bosons, $h_1$ and $h_2$, one of which is the SM-like 
Higgs with a mass of 125 GeV, and one DM scalar, which we call $A$.
Since the mixing between the two scalars is introduced only via the
rotation angle, the couplings of the two Higgs bosons to the remaining SM particles
is modified by the same factor $k_i$ defined as.
\begin{align}
\label{eq:HiggsVbosoncoupling}
g_{h_{i} SM \, SM}=g_{H_{\mathrm{SM}}SM \, SM} k_{i} \;, \; \;
  k_{i}\equiv \begin{cases}
\begin{array}{ll} \cos\alpha\;, & i=1 \\ -\sin\alpha \;, &i=2 \end{array}\end{cases},
\end{align}
where $g_{H_{\mathrm{SM}} SM \, SM}$ denotes the SM coupling between the SM Higgs and the SM particle $SM$.

With these definitions the parameters of the potential can now be
written as functions of our choice of input parameters given by
\begin{align}
v\,,\ v_{S}\,,\ \alpha\,,\ m_{h_1}\,,\ m_{h_2}\,,\ m_{A} \,,
\end{align}
as
\begin{subequations}
\label{eq:inputparameters}
\begin{align}
\lambda&= \frac{m_{h_1}^2+ m_{h_2}^2+ \cos2 \alpha (m_{h_1}^2-m_{h_2}^2)}{v^2}, \\
d_2&= \frac{m_{h_1}^2+ m_{h_2}^2+ \cos2 \alpha (m_{h_2}^2-m_{h_1}^2)}{v_{S}^2}, \\
\delta_2&= \frac{(m_{h_1}^2-m_{h_2}^2)\sin2 \alpha}{v v_S}, \\
m^2&=\frac{1}{2}\left( \cos2\alpha (m_{h_2}^2-m_{h_1}^2) -\frac{v(m_{h_1}^2+m_{h_2}^2)+v_S (m_{h_1}^2 -m_{h_2}^2)\sin2\alpha}{v} \right), \\
b_2&=\frac{1}{2}\left( 2 m_{A}^2 - m_{h_1}^2 -m_{h_2}^2 + \cos2\alpha (m_{h_1}^2-m_{h_2}^2) -\frac{v(m_{h_1}^2-m_{h_2}^2)\sin2\alpha}{v_S} \right), \\
b_1&=-m_{A}^{2} \,.
\end{align}
\end{subequations}
Note that the model depends only on 4 independent parameters, because
one of the Higgs bosons plays the role of the SM-like Higgs with a mass of 125 GeV
and the doublet VEV $v=1/\sqrt{\sqrt{2}G_F}\approx 246.22$~GeV, where
$G_F$ denotes the Fermi constant. The VEV is replaced by $G_F$ as an input parameter.

\section{Renormalization }
\label{ch:renormalization}

Our goal is to calculate the decay width of the Higgs bosons into a
pair of DM particles, $h_i \to AA$, at NLO. Since $A$ only couples to
the two Higgs bosons $h_i$ we just need to renormalize the scalar
sector. With the trilinear $h_i$ couplings to the DM particles given by
\beq
\lambda_{h_i AA} = \frac{m_{h_i}^2}{v_s} \begin{cases}
\begin{array}{ll} s_\alpha\,, & i =1 \\ c_\alpha \,, & i=2 \end{array}
\end{cases} ,
\label{eq:lhiAA}
\eeq
and according to our choice of input parameters we need to
renormalize the masses of the two
  scalars $h_i$, the mass of the DM particle, $m_A$, the singlet VEV $v_S$ and 
the mixing angle $\alpha$. 
Besides these parameters we also need to renormalize the $h_i$ and
  $A$ fields and the tadpoles to work with finite Green functions. We
start by formally defining the relation between the bare and the
renormalized quantities as  
\begin{align}
\label{eq:renormalizationquantity}
\beta_0=\beta+\delta \beta \, ,
\end{align}
where $\delta \beta$ is the counterterm of the physical quantity $\beta$ and $\beta_{0}$ is the bare quantity. All bare fields $\phi_0$ are related to their renormalized version via 
\begin{align}
\phi_{0}=\sqrt{Z_{\phi}}\phi\approx \left(1 + \frac{\delta Z_{\phi}}{2}\right) \phi \, ,
\end{align}
where $Z_{\phi}$ is the field strength renormalization constant. 

\subsection{On-Shell Renormalization of the Scalar Sector}
\label{sec:osscalarsector}

We start by calculating the mass and field counterterms in the scalar sector using the on-shell scheme. The renormalization constants for the DM particle are defined as
\begin{align}
\label{eq:renormalizationA}
A_{0}=\sqrt{Z_{A}}A \approx \left(1+ \frac{\delta Z_{A}}{2} \right) A, \; \quad D_{A,0}^{2}=D_{A}^{2}+ \delta D_{A}^{2},
\end{align}
where  $Z_A$ is the field strength renormalization constant, $D_{A,0}^{2} =m_{A,0}^{2}$  and $\delta D_A$ is the mass counterterm for $A$. 

The two scalars $h_1$ and $h_2$ again mix at one-loop order and
therefore both the field renormalization constants and the mass
counterterms are defined by 
\begin{align}
\label{eq:renormalizationhi}
\begin{pmatrix}
h_{1,0} \\
h_{2,0} \\
\end{pmatrix}=\sqrt{Z_{hh}}\begin{pmatrix}
h_{1} \\
h_{2} \\
\end{pmatrix} \approx \left(1 + \frac{\delta Z_{hh}}{2} \right) \begin{pmatrix}
h_{1} \\
h_{2} \\
\end{pmatrix}, \; D_{hh,0}^{2}=D_{hh}^{2}+ \delta D_{hh}^{2},
\end{align}
with $ D_{hh,0}^{2} = \mbox{diag}(m_{h1,0}^2, m_{h2,0}^2 )$ and the matrices $\delta Z_{hh}$ and $\delta D_{hh}^2$ defined as
\begin{align}
\label{eq:countertermmatrices}
\delta Z_{hh}= \begin{pmatrix}
\delta Z_{h_{1}h_{1}} & \delta Z_{h_{1}h_{2}}\\
\delta Z_{h_{2}h_{1}} & \delta Z_{h_{2}h_{2}} \\
\end{pmatrix}, \; \delta D_{hh}^{2}= \begin{pmatrix}
\delta D_{h_{1}h_{1}}^{2} & \delta D_{h_{1}h_{2}}^{2}\\
\delta D_{h_{1}h_{2}}^{2} & \delta D_{h_{2}h_{2}}^{2} \\
\end{pmatrix}.
\end{align}

The on-shell renormalization conditions lead to the following
expressions 
\begin{subequations}
\label{eq:ossolution}
\begin{align}
\delta D^2_{h_{i}h_{i}}&=\mathrm{Re} \left( \Sigma_{h_{i}h_{i}}(m_{h_{i}}^2) \right), \\
\delta Z_{h_{i}h_{i}}&= - \mathrm{Re} \left(\left.\frac{\partial \Sigma_{h_{i}h_{i}} (p^2)}{\partial p^2} \right|_{p^2=m^{2}_{h_{i}}} \right), \\
\delta Z_{h_{i}h_{j}}&=\frac{2}{m^2_{h_{i}}-m^2_{h_{j}}}\mathrm{Re}\left( \Sigma_{h_{i}h_{j}}(m_{h_{j}}^2) - \delta D^2_{h_{i}h_{j}}\right) (i \neq j),
\end{align}
\end{subequations}
for the counterterms of the scalar fields $h_i$ where
  $\Sigma_{h_i h_i}$ denotes their self-energies. Similarly, the
expressions for the DM field $A$ read  
\begin{subequations}
\label{eq:ossolutiona}
\begin{align}
\delta D^2_{A}&=\mathrm{Re} \left( \Sigma_{A}(m_{A}^2) \right), \\
\delta Z_{A}&= - \mathrm{Re} \left(\left.\frac{\partial \Sigma_{A} (p^2)}{\partial p^2} \right|_{p^2=m^{2}_{A}} \right).
\end{align}
\end{subequations}

The diagonal terms of $\delta D_{hh}^{2}$ or $\delta D_A^2$ are
related to the mass counterterms  and to the corresponding tadpoles.
 The off-diagonal terms are related to the tadpoles to be discussed in the next section.

\subsection{Tadpole Renormalization}
\label{sec:tadpolerenormalization}

Tadpole renormalization  is essentially the way we choose the VEVs at
1-loop order so that the minimum conditions hold. Another way to express it
is to state that the terms proportional to the scalar fields at 1-loop
order have to vanish. The VEV chosen to fulfil this condition~\cite{Fleischer:1980ub, deSousaMachadoFontes:2021zia} is the true VEV of the theory. 
We will follow the scheme proposed by Fleischer and Jegerlehner
\cite{Fleischer:1980ub} for the SM with the goal of rendering all
counterterms related to physical quantities gauge independent. The
scheme was applied to various extensions of the SM (see
e.g. \cite{Krause:2016oke,Krause:2017mal}). For the CxSM a brief description follows. 
We start by defining the true VEVs by performing the shifts
\begin{subequations}
\begin{align}
v &\rightarrow v + \Delta v, \\
v_{S} &\rightarrow v_{S}+ \Delta v_{S},
\end{align}
\end{subequations}
which lead to the following shifts in the tadpole parameters at NLO
\begin{subequations}
\begin{align}
T_{1} \rightarrow T_{1}+ \frac{v^2\lambda}{2}\Delta v + \frac{\delta_2vv_{S}}{2}\Delta v_{S} &\equiv T_1 +\delta T_{1}, \\
T_{2} \rightarrow T_{2}+ \frac{\delta_2vv_{S}}{2}\Delta v + \frac{d_2v_{S}^2}{2}\Delta v_{S} &\equiv T_2 +\delta T_{2}.
\end{align}
\end{subequations}
The minimum equations lead to the following relations between the
shifts in the VEVs and the tadpole counterterms 
\begin{align}
\label{eq:vevgaugetadpolemass}
\begin{pmatrix}
\Delta v \\ \Delta v_S \\
\end{pmatrix}=R_{\alpha}^{\mathrm{T}}\begin{pmatrix}
\frac{\delta T_{h_1}}{m_{h_1}^2} \\
\frac{\delta T_{h_2}}{m_{h_2}^2} 
\end{pmatrix} \;, 
\end{align}
with the relation between the tadpole counterterms
  $\delta T_{1,2}$ in the gauge basis and those in the mass basis, $\delta T_{h_{1,2}}$,
given by
\beq
\left( \begin{array}{c} \delta T_{1} \\ \delta T_{2} \end{array} \right)= R_\alpha^T
\left( \begin{array}{c} \delta T_{h_1} \\ \delta T_{h_2} \end{array}
\right) \;. \label{eq:rel2}
\eeq
The shift introduced in the VEVs can be applied to the mass matrix from Eq. \eqref{eq:massmatrix}. The additional terms resulting from that shift read
\begin{align}
\label{eq:Mshiftalternative}
\mathcal{M} \, \rightarrow \, \mathcal{M} + 
\begin{pmatrix}
v \Delta v \lambda & \frac{\delta_2}{2}(\Delta v v_S + v \Delta v_S) \\
\frac{\delta_2}{2}(\Delta v v_S + v \Delta v_S)  & d_2 v_S \Delta v_S \\
\end{pmatrix}
-
\underbrace{\begin{pmatrix}
\frac{T_1\Delta v}{v^2} & 0 \\
0 & \frac{T_2\Delta v_S}{v_S^2}
\end{pmatrix}}_{\text{vanishes}}.
\end{align}
The last term in Eq. \eqref{eq:Mshiftalternative} vanishes, because after the shift the tadpole conditions can be applied again. The mass matrix can now be rotated into the mass basis and all counterterm shifts can be applied leading to
\begin{align}
\begin{split}
D_{hh}^{2}=R_{\alpha}\mathcal{M}R_{\alpha}^{\mathrm{T}} \, &\rightarrow  \, D_{hh}^{2} + 
\begin{pmatrix}
\delta m_{h_1}^2 & 0 \\
0 & \delta m_{h_2}^2 \\
\end{pmatrix} \\
&+
R_{\alpha}
\begin{pmatrix}
\frac{\delta T_1}{v} + v \Delta v \lambda & \frac{\delta_2}{2}(\Delta v v_S + v \Delta v_S)\\
\frac{\delta_2}{2}(\Delta v v_S + v \Delta v_S) & \frac{\delta T_2}{v_S}  + d_2 v_S \Delta v_S \\
\end{pmatrix}
R_{\alpha}^{\mathrm{T}}\\ 
&\equiv D_{hh}^{2} + \begin{pmatrix} \delta m_{h_1}^2 & 0 \\
0 & \delta m_{h_2}^2 \\
\end{pmatrix} + 
\begin{pmatrix}
\Delta D_{h_1h_1}^2 & \Delta D_{h_1h_2}^2 \\
\Delta D_{h_1h_2}^2 & \Delta D_{h_2h_2}^2 \\
\end{pmatrix} . 
\end{split}
\end{align}
Using Eqs.~(\ref{eq:vevgaugetadpolemass}) and (\ref{eq:rel2}) as well
  as the relations Eq.~(\ref{eq:inputparameters}) between the potential
  parameters and the input parameters we can express the shifts
  $\Delta D^2_{h_i h_j}$ ($i,j=1,2$) as
\begin{subequations}
\begin{align}
\Delta D_{h_1h_1}^{2}&= i(-i\lambda_{h_1h_1h_1}) \frac{-i}{m_{h_1}^2} i\delta T_{h_1} + i(-i\lambda_{h_1h_1h_2}) \frac{-i}{m_{h_2}^2} i\delta T_{h_2}, \\
\Delta D_{h_1h_2}^{2}&= i(-i\lambda_{h_1h_1h_2}) \frac{-i}{m_{h_1}^2} i\delta T_{h_1} + i(-i\lambda_{h_1h_2h_2}) \frac{-i}{m_{h_2}^2} i\delta T_{h_2}, \\
\Delta D_{h_2h_2}^{2}&= i(-i\lambda_{h_1h_2h_2}) \frac{-i}{m_{h_1}^2}
                       i\delta T_{h_1} + i(-i\lambda_{h_2h_2h_2})
                       \frac{-i}{m_{h_2}^2} i\delta T_{h_2},
\end{align}
\end{subequations}
with the trilinear Higgs couplings given by
\begin{subequations}
\begin{align}
\lambda_{h_1h_1h_1}&=3m_{h_1}^2\frac{v_S c_{\alpha}^3 + v s_{\alpha}^3}{v v_S}, \\
\lambda_{h_1h_1h_2}&= \frac{(2m_{h_1}^2+ m_{h_2}^2)s_{\alpha} c_{\alpha}(v s_{\alpha}- v_S c_{\alpha})}{v v_S},\\
\lambda_{h_1h_2h_2}&= \frac{(m_{h_1}^2+ 2 m_{h_2}^2)s_{\alpha} c_{\alpha}(v c_{\alpha} + v_S s_{\alpha})}{v v_S},\\
\lambda_{h_2h_2h_2}&=3m_{h_2}^2\frac{ v c_{\alpha}^3 -v_S s_{\alpha}^3}{v v_S}.
\end{align}
\end{subequations}
In terms of Feynman diagrams this can be seen as the contribution of the tadpole diagram (times a factor $i$, at vanishing momentum transfer) to the propagators of $h_1$ and $h_2$, 
which were not included previously in the definition of the self-energies. We define
\begin{align}
i\Sigma_{h_{i}h_{j}}^{\mathrm{tad}}(p^2) \equiv i\Sigma_{h_{i}h_{j}}(p^2) - i\Delta D_{h_ih_j}^2,
\end{align}
and the renormalized self-energies take the form
\begin{align}
\begin{split}
\hat{\Sigma}_{h_{i}h_{j}}(p^2)=\Sigma_{h_{i}h_{j}}^{\mathrm{tad}}(p^2)-\begin{pmatrix}
\delta m_{h_1}^2 & 0 \\
0 & \delta m_{h_2}^2 \\
\end{pmatrix}
 &+ \frac{\delta Z_{h_{i}h_{j}}^{\dagger}}{2}\left( p^2 \delta_{h_ih_j}-D_{h_{i}h_{j}}^2\right) \\
 &+ \left(p^2 \delta_{h_ih_j}-D_{h_{i}h_{j}}^2 \right)\frac{\delta Z_{h_{i}h_{j}}}{2}.
 \end{split}
\end{align}
This shift of contributions from the mass counterterm matrix into the self-energy corresponds to the inclusion of the tadpole diagrams into the self-energy. 
With this change in the renormalized self-energy the following results for the counterterms hold
\begin{subequations}
\label{eq:solutionalternativescheme}
\begin{align}
\delta m_{h_{i}}^2&=\mathrm{Re} \left( \Sigma_{h_{i}h_{i}}^{\mathrm{tad}}(m_{h_{i}}^2) \right), \\
\delta Z_{h_{i}h_{i}}&= - \mathrm{Re} \left(\left.\frac{\partial \Sigma_{h_{i}h_{i}}^{\mathrm{tad}} (p^2)}{\partial p^2} \right|_{p^2=m^{2}_{h_{i}}} \right), \\
\delta Z_{h_{i}h_{j}}&=\frac{2}{m^2_{h_{i}}-m^2_{h_{j}}}\mathrm{Re}\left( \Sigma_{h_{i}h_{j}}^{\mathrm{tad}}(m_{h_{j}}^2)\right) (i \neq j).
\end{align}
\end{subequations}

Following a similar reasoning, the counterterms of the field $A$ can be expressed as
\begin{align}
\delta m_{A}^2&=\mathrm{Re} \left( \Sigma_{A}^{\mathrm{tad}}(m_{A}^2) \right), \\
\delta Z_{A}&= - \mathrm{Re} \left(\left.\frac{\partial \Sigma_{A}^{\mathrm{tad}} (p^2)}{\partial p^2} \right|_{p^2=m^{2}_{A}} \right).
\end{align}

\subsection{Renormalization of the Mixing Angle $\alpha$}
\label{sec:renormalizationofalpha}
There are two parameters left to be renormalized. We start with the
rotation angle $\alpha$.  Previous works~\cite{Krause:2016oke,
  Azevedo:2021ylf} lead us to the conclusion that a scheme that is
simultaneously stable (in the sense that the NLO corrections do not
become unreasonably large) and gauge independent
can be built by combining the one proposed
in~Ref.~\cite{Kanemura:2004mg, Pilaftsis:1997dr} with the gauge
dependence handled by the use of the pinch
technique~\cite{Cornwall:1989gv, Papavassiliou:1994pr}.   
The scheme proposed in~\cite{Kanemura:2004mg, Pilaftsis:1997dr}
introduces a shift in $\alpha$, the angle from the rotation matrix
$R_\alpha$,  
\begin{align}
\label{eq:mixingmatrixcounterterm}
R_{\alpha,0} \approx R_{\delta \alpha}R_{\alpha},
\end{align}
and by relating it to the field renormalization matrix constant leads
to the following counterterm for $\alpha$, 
\begin{align}
\delta \alpha = \frac{\delta Z_{h_1h_2}- \delta Z_{h_2h_1}}{4}.
\end{align}
The result is model independent, it only assumes
  the mixing of solely two fields.
This relation can now be expressed in terms of self-energies as
\begin{align}
\label{eq:Kosyresult}
\delta \alpha = \frac{1}{2(m_{h_{1}}^2-m_{h_2}^2)}\mathrm{Re}\left(
  \Sigma^{\text{tad}}_{h_1h_2}(m_{h_1}^2)+\Sigma^{\text{tad}}_{h_1h_2}(m_{h_2}^2)
  \right). 
\end{align}
This counterterm turns out to be gauge dependent. This in itself would
not be a problem if the complete amplitude for the process was gauge
independent, which is not the case. There is, however,   
a procedure to isolate this gauge dependence in a systematic and
consistent way known as the pinch
technique~\cite{Papavassiliou:1989zd, Cornwall:1989gv,
  Papavassiliou:1994pr, Binosi:2009qm}. 
After successfully applying the pinch technique, the pinched
self-energies can be defined by adding the additional contributions to
the self-energies from the pinch technique. This results in 
\begin{align}
\label{eq:pinchedselfenergies}
\begin{split}
i\Sigma^{\text{pinch}}_{h_{i}h_{j}}(p^2)&=i\Sigma^{\text{tad}}_{h_{i}h_{j}}(p^2) + i\Sigma^{\text{add}}_{h_{i}h_{j}}(p^2)\\
&=\left.i\Sigma^{\text{tad}}_{h_{i}h_{j}}(p^2)\right|_{\{\xi =1\}} \\
&+ \frac{-ig^2}{32\pi^2c_{\mathrm{w}}^2} \left( p^2-\frac{m_{h_i}^2+m_{h_{j}}^2}{2} \right) O_{ij}B_{0}(q^2,m_{Z}^2,m_{Z}^2) \\
&+ \frac{-ig^2}{16\pi^2} \left( p^2-\frac{m_{h_i}^2+m_{h_j}^2}{2} \right) O_{ij}B_{0}(q^2,m_{W}^2,m_{W}^2).
\end{split}
\end{align}
The loop integral $B_0$ and the factor $O_{ij}$ as well as
$\Sigma^{\text{add}}_{h_{i}h_{j}}(p^2)$ are defined in
App.~\ref{sec:pinchtechnique}. Note that the expression with $\xi=1$
does not mean that a specific gauge has been chosen. The additional terms 
together with the tadpole self-energies result in a gauge-independent result which can just be written in that form. We can now define a gauge-independent counterterm for $\alpha$, for which two different scales will be chosen:
\begin{itemize}
\item Setting the external momenta to the respective OS masses, $p^2=m_{h_{i}}^2 $, called OS pinched scheme.
\item Setting the external momenta to the mean of the masses, $p^2=p^2_{*}=\frac{m_{h_{1}}^2+m_{h_{2}}^2}{2}$, called $p^{*}$ pinched scheme.
\end{itemize}
  In the $p^{*}$ pinched scheme the additional gauge-independent terms
  from the pinch technique vanish so that the expression for the
  mixing angle counterterm becomes more compact. 
  We
can write the counterterm  for $\alpha$ in the $p_{*}$ scheme and the
OS pinched scheme as 
\begin{subequations}
\begin{align}
\delta \alpha_{p^*} &= \frac{1}{(m_{h_{1}}^2-m_{h_2}^2)}\mathrm{Re}\left( \left.\Sigma^{\text{tad}}_{h_1h_2}(p^2_{*})\right|_{\{\xi=1\}}\right),\\
\delta \alpha_{\text{OS}}&=\frac{1}{2(m_{h_{1}}^2-m_{h_2}^2)} \mathrm{Re}\left(  \Sigma^{\text{pinch}}_{h_1h_2}(m_{h_1}^2)+\Sigma^{\text{pinch}}_{h_1h_2}(m_{h_2}^2) \right).
\end{align}
\end{subequations}
With these definitions, $\delta \alpha$ is gauge independent by
construction and the problem with the gauge dependence is solved. 

\subsection{Renormalization of $v_S$}
\label{sec:renormalizationvs}
The last parameter to be renormalized is the VEV $v_S$ of the scalar singlet.
We will be using a process-dependent scheme and also a derivation
thereof where the conditions are imposed at the amplitude and not at
the physical process level, defined as zero external momentum scheme
(ZEM) scheme~\cite{Azevedo:2021ylf}.
The latter, although less stable, allows to cover the entire parameter
space because it is not constrained by kinematic restrictions. 

\subsubsection{Process-dependent Scheme}
\label{sec:processdependentscheme}
The process to be used needs a coupling constant proportional to
$v_S$ and if we want to use a decay,  the only possibilities\footnote{In principle the decay $h_2\rightarrow h_1h_1$ could also be chosen but would lead to an additional kinematic constraint between the two scalar masses $m_{h_1}$ and $m_{h_{2}}$ and would constrain the parameter space even more.} in the
CxSM are  $h_1 \rightarrow AA$ and $h_2 \rightarrow AA$. Therefore one
of the processes will be used 
to extract the singlet VEV renormalization constant, and because we
want to use the measurement of the SM-like Higgs invisible width, the
second Higgs will be used for that purpose. Note, however, that any of
the two Higgs bosons can be the SM-like one, while the other can
either be lighter or heavier than 125 GeV. Hence, there are two
scenarios to be analysed and we have to find $v_S$ for both.   

In the process-dependent scheme the counterterm is calculated by forcing
\begin{align}
\label{eq:processdependentschemegeneralequation}
\Gamma^{\mathrm{LO}}_{h_i \rightarrow AA}=\Gamma^{\mathrm{NLO}}_{h_i\rightarrow AA},
\end{align}
that is, the LO and NLO decay widths are equal. This is turn leads to
\begin{align}
\label{eq:pdssimplified}
0=\mathrm{Re}\left( \left( \mathcal{A}^{\mathrm{LO}}_{h_i\rightarrow AA} \right)^*  \mathcal{A}^{\mathrm{NLO}}_{h_i\rightarrow AA} \right),
\end{align}
where $\mathcal{A}^{\mathrm{LO}}_{h_i\rightarrow AA}$ is the amplitude of the process $h_i \rightarrow AA$ at LO and $\mathcal{A}^{\mathrm{NLO}}_{h_i\rightarrow AA}$ is the amplitude at NLO. Because the LO amplitude
is just a coupling constant, the expression further simplifies to
\begin{align}
\label{eq:pdssimplifiedfurther}
0=\mathrm{Re} \left( \mathcal{A}^{\mathrm{NLO}}_{h_i\rightarrow AA} \right).
\end{align}
The NLO contribution $\mathcal{A}^{\mathrm{NLO}}_{h_i\rightarrow AA}$ can be written in terms of the vertex corrections $\mathcal{A}^{\mathrm{VC}}_{h_i\rightarrow AA}$ and the vertex counterterm such that
\begin{align}
\begin{split}
0&=\mathrm{Re} \left( \mathcal{A}^{\mathrm{NLO}}_{h_i\rightarrow AA} \right)\\
&= \mathrm{Re}\left( \mathcal{A}^{\text{VC}}_{h_i\rightarrow AA} \right) - \lambda_{h_i AA} \left( \frac{\delta \lambda_{h_i AA}}{\lambda_{h_i AA}} + \delta Z_A + \frac{\delta Z_{h_ih_i}}{2} + \frac{\lambda_{h_j AA}}{\lambda_{h_i AA}}\frac{\delta Z_{h_jh_i}}{2} \right),
\end{split}
\end{align}
where $i,j \in \{1,2\}$, but $i \neq j$. And with 
  the trilinear $h_i$ couplings to the DM particles $\lambda_{h_i AA}$
  given in Eq.~(\ref{eq:lhiAA})
we have
\begin{equation}
\label{eq:couplingcounterterm}
\frac{\delta \lambda_{h_iAA}}{\lambda_{h_iAA}} = \frac{\delta
  m_{h_{i}}^2}{m_{h_{i}}^2} - \frac{\delta v_S}{v_S} + T_{i}(\alpha)
\delta \alpha, \quad T_{i}(\alpha)
\equiv \begin{cases} \begin{array}{ll} \cot \alpha\,, & i=1 \\ -\tan
                       \alpha\,, & i=2 \end{array}\end{cases}.
\end{equation}
Finally, the expression for the counterterm $v_S$ reads  
\begin{subequations}
\begin{align}
\begin{split}
\delta v_S^{h_1 \rightarrow AA}= v_S \left( -\mathrm{Re} \left( \frac{\mathcal{A}^{\text{VC}}_{h_1\rightarrow AA} }{\lambda_{h_1AA}} \right)+ \frac{\delta m_{h_1}^2}{m_{h_1}^{2}} \right. &+\cot \alpha \, \delta \alpha +  \delta Z_A  \\
&+ \left. \frac{\delta Z_{h_1h_1}}{2} + \frac{\lambda_{h_2AA}}{\lambda_{h_1AA}} \frac{\delta Z_{h_2h_1}}{2} \right),
\end{split} \\
\begin{split}
\delta v_S^{h_2 \rightarrow AA}= v_S \left( -\mathrm{Re} \left( \frac{ \mathcal{A}^{\text{VC}}_{h_2\rightarrow AA} }{\lambda_{h_2AA}} \right) + \frac{\delta m_{h_2}^2}{m_{h_2}^{2}} \right. &-\tan \alpha  \, \delta \alpha +  \delta Z_A  \\
&+ \left.\frac{\delta Z_{h_2h_2}}{2} + \frac{\lambda_{h_1AA}}{\lambda_{h_2AA}} \frac{\delta Z_{h_1h_2}}{2} \right),
\end{split}
\end{align}
\end{subequations}
for the two processes. These counterterms are gauge independent and
lead to UV-finite results. The renormalization scheme also leads to
stable results. Therefore, the only drawback is the kinematic restriction
\begin{align}
m_{h_i}> 2 m_A,
\end{align}
which forces us to be in a restricted region of the parameter space. We discuss a solution to avoid this restriction in the next section.

\subsubsection{ZEM Scheme}
\label{sec:zemscheme}
The ZEM scheme was introduced in \cite{Azevedo:2021ylf} to avoid
kinematic restrictions on the parameter space, and we will now apply
it to the CxSM. It is a simple derivation of the process-dependent
scheme, where the square of all external momenta are set to zero at
the level of the amplitude, 
\begin{align}
p_{\mathrm{in}}^{2}=p_{\mathrm{out}1}^2=p_{\mathrm{out}2}^2=0,
\end{align}
eliminating therefore the kinematic constraint. Choosing the same
physical processes, the condition now reads 
\begin{align}
\label{eq:ZEMStart}
0=\mathrm{Re} \left( \mathcal{A}^{\mathrm{NLO}}_{h_i\rightarrow AA}(\{p^2=0\}) \right),
\end{align}
where $p^2=0$ means that all squared external momenta are set to zero. There is another difference relative to the process-dependent scheme: the NLO leg corrections are not canceled by the corresponding counterterms, 
because the leg counterterms are defined through the OS
scheme. Therefore  Eq.~(\ref{eq:ZEMStart}) now takes the form
\begin{align}
\begin{split}
0&=\mathrm{Re} \left( \mathcal{A}_{\mathrm{VC}}^{h_i \rightarrow AA}(\{p^2=0\}) + \mathcal{A}_{\mathrm{Leg}}^{h_i \rightarrow AA}(\{p^2=0\}) \right),\\
&+\lambda_{h_iAA}\left( -\frac{\delta \lambda_{h_iAA}}{\lambda_{h_iAA}} + \delta Z_{A} + \frac{\delta Z_{h_{i}h_{i}}}{2} + \frac{\delta m_{h_{i}}^{2}}{m_{h_{i}}^{2}} +  \frac{2\delta m_{A}^{2}}{m_{A}^{2}}  +\frac{\lambda_{h_jAA}}{\lambda_{h_iAA}} \frac{m_{h_{i}}^{2}}{m_{h_{j}}^{2}} \frac{\delta Z_{h_{i}h_{j}}}{2} \right).
\end{split}
\end{align}
Again, this equation can be solved for the two processes $h_1 \rightarrow AA$ and $h_2 \rightarrow AA$ to obtain the counterterms 
\begin{subequations}
\label{eq:ZEMsol}
\begin{align}
\begin{split}
\delta v_{S}^{\mathrm{ZEM}, h_1\rightarrow AA}&=v_S \left( -\mathrm{Re} \left( \frac{\mathcal{A}_{\mathrm{VC}}^{h_1 \rightarrow AA}(\{p^2=0 \})+\mathcal{A}_{\mathrm{Leg}}^{h_1 \rightarrow AA}(\{p^2=0 \})}{\lambda_{h_1AA}} \right) \right. \\
& \left. +\cot \alpha \, \delta \alpha -\delta Z_{A}-\frac{2\delta m_{A}^{2}}{m_A^2}-\frac{\delta Z_{h_1 h_1}}{2}-\cot\alpha\frac{\delta Z_{h_1 h_2}}{2} \right)
\end{split} \\
\begin{split}
\delta v_{S}^{\mathrm{ZEM}, h_2\rightarrow AA}&=v_S \left( -\mathrm{Re} \left( \frac{\mathcal{A}_{\mathrm{VC}}^{h_2 \rightarrow AA}(\{p^2=0 \})+\mathcal{A}_{\mathrm{Leg}}^{h_2 \rightarrow AA}(\{p^2=0 \})}{\lambda_{h_2AA}} \right) \right. \\
&\left. -\tan\alpha \, \delta \alpha -\delta Z_{A}-\frac{2\delta m_{A}^{2}}{m_A^2}-\frac{\delta Z_{h_2 h_2}}{2}-\tan\alpha \frac{\delta Z_{h_2 h_1}}{2} \right).
\end{split}
\end{align}
\end{subequations}
We now just have to check if the final result is finite and gauge independent. The question of gauge dependence in the alternative tadpole scheme is always related to wave function renormalization
constants. A thorough analysis leads to the conclusion that although finite the result is gauge dependent due to the term
\begin{align}
\frac{\delta Z_{h_{i}h_{i}}}{2}  +\frac{\lambda_{j}}{\lambda_{i}} \frac{m_{h_{i}}^{2}}{m_{h_{j}}^{2}} \frac{\delta Z_{h_{i}h_{j}}}{2},
\end{align}
for the corresponding process $h_i \rightarrow AA$. The problem was solved  by simply replacing the self-energies in the wave function renormalization constants in Eq.~(\ref{eq:ZEMsol}) by their pinched versions. 
This way $\delta v_S$ becomes gauge independent. This change in the
$\delta Z_{h_ih_j}$, however, is only applied to terms appearing in
Eq.~\eqref{eq:ZEMsol} where the ZEM counterterm of $v_S$ is defined
and not anywhere else.  
Otherwise, a gauge dependence in the overall amplitude of the
renormalized process could be reintroduced.  
Therefore, the resulting counterterms for $v_S$ in this modified ZEM scheme read
\begin{subequations}
\label{eq:ZEMsolGI}
\begin{align}
\begin{split}
\delta v_{S}^{\mathrm{ZEMGI}, h_1\rightarrow AA}&=v_S \left( -\mathrm{Re} \left( \frac{\mathcal{A}_{\mathrm{VC}}^{h_1 \rightarrow AA}(\{p^2=0 \})+\mathcal{A}_{\mathrm{Leg}}^{h_1 \rightarrow AA}(\{p^2=0 \})}{\lambda_{h_1AA}} \right) \right. \\
& \left. + \cot \alpha \, \delta \alpha -\delta Z_{A}-\frac{2\delta m_{A}^{2}}{m_A^2}-\frac{\delta Z_{h_1 h_1}^{\text{pinched}}}{2}-\tan \alpha \, \frac{\delta Z_{h_1 h_2}^{\text{pinched}}}{2} \right)
\end{split}\\
\begin{split}
\delta v_{S}^{\mathrm{ZEMGI}, h_2\rightarrow AA}&=v_S \left( - \mathrm{Re} \left( \frac{\mathcal{A}_{\mathrm{VC}}^{h_2 \rightarrow AA}(\{p^2=0 \})+\mathcal{A}_{\mathrm{Leg}}^{h_2 \rightarrow AA}(\{p^2=0 \})}{\lambda_{h_2AA}} \right) \right. \\
&\left. -\tan \alpha \, \delta \alpha -\delta Z_{A}-\frac{2\delta m_{A}^{2}}{m_A^2}-\frac{\delta Z_{h_2 h_2}^{\text{pinched}}}{2}-\tan \alpha \, \frac{\delta Z_{h_2 h_1}^{\text{pinched}}}{2} \right).
\end{split}
\end{align}
\end{subequations}

 The renormalization is now complete and before moving to the presentation of the NLO results we will discuss the constraints imposed on the model.
  
\section{Constraints on the Model}
\label{sec:constraints}

The constraints imposed to find the allowed parameter space are
implemented in \texttt{ScannerS}~\cite{Coimbra:2013qq,
  PhysRevD.92.025024, Muhlleitner:2020wwk}. 
In this section we will just briefly review the most relevant
theoretical and experimental constraints considered.  

\subsection{Theoretical Constraints}
\label{sec:theoreticalconstraints}
\begin{itemize}

\item \textbf{Boundedness from Below}

The conditions to have a stable minimum are easily obtained by
writing, $\Phi^{\dagger}\Phi \equiv x$ and $|\mathbb{S}|^2 \equiv y$
and writing the quartic terms of the potential 
\begin{align}
V_{\mathrm{quartic}}(x,y)= \frac{\lambda}{4}x^2 + \frac{\delta_2}{2}xy + \frac{d_2}{4}y^2 = \frac{1}{4} \begin{pmatrix}
x & y 
\end{pmatrix}^{\mathrm{T}} \begin{pmatrix}
\lambda & \delta_2 \\
\delta_2 & d_2 
\end{pmatrix}\begin{pmatrix}
x \\ 
y 
\end{pmatrix}.
\end{align}
Forcing the potential to be bounded in all
directions leads to the 
following conditions at tree level 
\begin{align}
\label{eq:boundedfrombelow}
\lambda >0 \wedge d_2 > 0 \wedge (\delta_2^2 < \lambda d_2 \, \mathrm{if} \, \delta_2 <0).
\end{align}
\item \textbf{Perturbative Unitarity Constraints} 

Following~\cite{Lee:1977eg} we force the eigenvalues of the scattering matrix $\mathcal{M}_{2 \rightarrow 2}$ of all possible two-to-two scalar scattering interactions to obey
\begin{align}
|\lambda_i| < 8 \pi \, ,
\end{align}
leading to
\begin{align}
\label{eq:unitarity}
\begin{split}
|\lambda|\leq 16\pi \wedge |d_2| \leq 16\pi \wedge |\delta_2| &\leq 16\pi \\ 
\wedge \left|\frac{3}{2}\lambda + d_2 \pm \sqrt{\left(\frac{3}{2}\lambda - d_2\right)^2+2\delta_2^2}\right| &\leq 16\pi.
\end{split}
\end{align}

\item \textbf{Stability of the Vacuum}

In the CxSM the most general vacuum structure is obtained by the following expectation values for the fields
\begin{align}
 \left<  \Phi \right> = \frac{1}{\sqrt{2}}
\begin{pmatrix}
0 \\
v
\end{pmatrix} , \; \; \left< \mathbb{S} \right> =\frac{1}{\sqrt{2}}\left(v_S + i v_A \right),
\end{align}
because of the $SU(2)$ invariance. Therefore, the value of the tree-level potential at each vacuum configuration is given by $V (v, v_S, v_A)$. We have chosen to work in the
configuration where the potential is  $V (v, v_S, 0)$ to have one DM
candidate. In  App.~\ref{sec:minima} we show that by choosing the vacuum
configuration with non-zero $v$ and $v_S$ (and $v_A$ =0)  to be a minimum automatically implies that this configuration is the absolute minimum at tree level.
\end{itemize}
%
%
%
\subsection{Experimental Constraints}
Before moving to the experimental constraints we note that $\rho=
m_W^2/(m_Z^2 c^2_{\mathrm{w}})$ where $m_{W,Z}$ are
  the masses of the massive $W$ and $Z$ bosons, respectively, and
  $c_{\mathrm{w}}$ denotes the cosine of the Weinberg angle, is equal
to $1$ at tree-level, like in the SM. 
Also, no tree-level flavour-changing neutral currents are introduced because the
gauge singlet does not couple to fermions and to gauge bosons in the
gauge basis.   

We will now briefly review the experimental constraints implemented in \texttt{ScannerS} and used for the generation of parameter points. 
\begin{itemize}
\item \textbf{$S,T,U$  precision parameters} \\
The additional scalar fields in the CxSM contribute to the gauge bosons self-energies and this implies deviations from the SM predictions. 
These deviations relative to the SM have to be within experimental
bounds, {\it i.e.}~\texttt{ScannerS} compares the model predictions with the
electroweak precision results from experiment.  
Then the program applies a consistency check on the
  $S,T,U$ parameters~\cite{PhysRevD.46.381} with 95 \% confidence
level to check if the constraints are fulfilled. 
 
\item  \textbf{Compatibility with the LHC Higgs data and exclusion bounds}\\
There are two important constraints coming from colliders. The most
relevant one is the one coming from the LHC related 
to the   measurements of the discovered Higgs
  boson. The searches for additional scalars also play a role in restricting the parameter space of the model.
\texttt{ScannerS} enforces these bounds by the interfaces with
\texttt{HiggsSignals} \cite{Bechtle:2013xfa, Bechtle:2020uwn} and
\texttt{HiggsBounds} \cite{Bechtle:2008jh,
  Bechtle:2020pkv}. Agreement of the signal rates of the
SM-like Higgs boson of the CxSM with the observations at $2 \sigma$
level is checked by \texttt{HiggsSignals}-2.6.1. Through
\texttt{HiggsBounds}-5.9.0 the exclusion bounds from searches for
extra scalars are taken into account. 

\item  \textbf{DM relic density}\\
The CxSM has a scalar  DM candidate and therefore the predicted DM relic density of this model should not exceed the measured value. Smaller values are not excluded since they allow for additional contributions coming from other sources.
 \texttt{ScannerS} is interfaced with the program package \texttt{MicrOMEGAs}~\cite{Belanger:2006is} to include this constraint from the relic density. 

\item  \textbf{DM direct detection}\\
As previously stated, the DM-nucleon cross section is only relevant at
one-loop order due to a cancellation that renders the tree-level cross
section proportional to the DM velocity and therefore
negligible~\cite{Gross:2017dan,Azevedo:2018oxv}.  
However, one-loop corrections to the DM-nucleon spin-independent cross
section have to be below the present experimentally measured result
from XENON1T~\cite{XENON:2018voc}, as discussed
in~\cite{Azevedo:2018exj, Glaus:2020ihj}. We will come 
back to this important constraint in the next section.   
\end{itemize}
  
\section{Results and Discussion} 
\label{sec:resdis}

\subsection{Higgs Decay into Dark Matter}
\label{ch:calculation}

The CxSM has two CP-even scalars $h_1$ or $h_2$ and any of them can
play the role of the 125 GeV SM-like Higgs boson
  denoted $h_{125}$ in the following. The non SM-like
Higgs can be either heavier or lighter than 125 GeV. In order to
optimize the analysis we fixed  
$h_1$ to always be the lightest of the two and considered two distinct scenarios,
\begin{itemize}
\item $m_{h_{1}}=m_{h_{125}}$ (scenario I): the width is calculated from $h_1 \rightarrow AA$ and the process $h_2 \rightarrow AA$ is chosen for the renormalization of $v_S$.
\item $m_{h_{2}}=m_{h_{125}}$ (scenario II): the width is calculated from $h_2 \rightarrow AA$ and the process $h_1 \rightarrow AA$ is chosen for the renormalization of $v_S$.
\end{itemize}

We now proceed to the calculation of the 125 GeV Higgs partial decay
width into two DM particles at electroweak NLO.
The calculations of the NLO corrections were performed using {\tt FeynRules~2.3.35}~\cite{Christensen:2008py,Degrande:2011ua,Alloul:2013bka},
 {\tt FeynArts~3.10}~\cite{Kublbeck:1990xc,Hahn:2000kx} and {\tt FeynCalc~9.3.1}~\cite{Mertig:1990an,Shtabovenko:2016sxi}. 
 Loop integrals were computed using LoopTools~\cite{Hahn:1998yk, vanOldenborgh:1989wn}.
 The model file was independently generated using  {\tt
   SARAH~4.14.2}~\cite{Staub:2009bi,Staub:2010jh,Staub:2012pb,Staub:2013tta,Staub:2015kfa}. We
 performed two independent calculations and found agreement between
 both results.

The LO decay width is given by
\begin{align}
\label{eq:decaywidthLO}
\Gamma^{\mathrm{LO}}_{h_{125} \rightarrow AA}=\frac{\lambda(m_{h_{125}}^2,m_A^2,m_A^2)}{32\pi m_{h_{125}}^3}\left| \mathcal{A}^{\mathrm{LO}}_{h_{125} \rightarrow AA} \right|^2.
\end{align}
while the NLO expression can be written as 
\begin{align}
\label{eq:decaywidthNLOhAA}
\Gamma^{\mathrm{NLO}}_{h_{125} \rightarrow AA}=\frac{\lambda(m_{h_{125}}^2,m_A^2,m_A^2)}{32\pi m_{h_{125}}^3}\left(\left| \mathcal{A}^{\mathrm{LO}}_{h_{125} \rightarrow AA} \right|^2 + 2\mathrm{Re}\left( \left(\mathcal{A}^{\mathrm{LO}}_{h_{125} \rightarrow AA}\right)^{*} \mathcal{A}^{\mathrm{NLO}}_{h_{125} \rightarrow AA}\right) \right),
\end{align}
with $\lambda (x, y, z) = x^2 + y^2 + z^2 - 2 xy - 2 xz - 2 yz$ and
$\mathcal{A}^{\mathrm{LO}}$ and $\mathcal{A}^{\mathrm{NLO}}$ denoting
the LO and NLO amplitudes, respectively. 

The LO amplitude is simply the coupling constant
\begin{align}
i\mathcal{A}^{\mathrm{LO}}_{h_{i} \rightarrow AA}= -i\lambda_{h_i AA}.
\end{align} 
and therefore the decay width takes the form 
\begin{subequations}
\label{eq:LOresult}
\begin{align}
\Gamma^{\mathrm{LO}}_{h_{1} \rightarrow AA}&=\frac{s_{\alpha}^{2}  m_{h_{1}} \lambda(m_{h_{1}}^2,m_A^2,m_A^2)}{32\pi v_S^{2}}, \\
\Gamma^{\mathrm{LO}}_{h_{2} \rightarrow AA}&=\frac{c_{\alpha}^{2}  m_{h_{2}} \lambda(m_{h_{2}}^2,m_A^2,m_A^2)}{32\pi v_S^{2}},
\end{align}
\end{subequations}
where both $h_1$ and $h_2$ can be the SM-like Higgs $h_{125}$.

For the NLO amplitude we need to compute the vertex corrections
together with the counterterm contributions. The vertex corrections
are just the sum of all irreducible contributions at 1-loop order
while  the vertex counterterm can be read off the Lagrangian yielding 
\begin{align}
\label{eq:vertexcounterterm}
\mathcal{A}_{h_i \rightarrow AA}^{\mathrm{CT}}=-\lambda_{h_iAA}\left( \frac{\delta \lambda_{h_iAA}}{\lambda_{h_iAA}} + \delta Z_{A} + \frac{\delta Z_{h_{i}h_{i}}}{2} + \frac{\lambda_{h_jAA}}{\lambda_{h_iAA}} \frac{\delta Z_{h_{j}h_{i}}}{2} \right),
\end{align}
where $i,j \in \{1,2\}$ but $i \neq j$. We finally arrive at the overall NLO contributions for the processes $h_i \rightarrow AA$
\begin{subequations}
\label{eq:nloamplitudehiAA}
\begin{align}
\begin{split}
\mathcal{A}_{h_1 \rightarrow AA}^{\mathrm{NLO}}= \mathcal{A}_{h_1 \rightarrow AA}^{\mathrm{VC}} -\lambda_{h_1AA}\left( \frac{\delta m_{h_1}^{2}}{m_{h_1}^2} \right. & - \frac{\delta v_S}{v_S}+ \cot \alpha \, \delta \alpha \\ 
+&\left.  \delta Z_{A} + \frac{\delta Z_{h_{1}h_{1}}}{2} + \cot \alpha \frac{m_{h_2}^2}{m_{h_1}^2} \frac{\delta Z_{h_{2}h_{1}}}{2} \right),
\end{split}\\
\begin{split}
\mathcal{A}_{h_2 \rightarrow AA}^{\mathrm{NLO}}= \mathcal{A}_{h_2 \rightarrow AA}^{\mathrm{VC}} -\lambda_{h_2AA}\left( \frac{\delta m_{h_2}^{2}}{m_{h_2}^2} \right. & - \frac{\delta v_S}{v_S}+ \cot \alpha \, \delta \alpha  \\
+&\left.  \delta Z_{A} + \frac{\delta Z_{h_{2}h_{2}}}{2} + \tan \alpha \frac{m_{h_1}^2}{m_{h_2}^2} \frac{\delta Z_{h_{1}h_{2}}}{2} \right).
\end{split}
\end{align}
\end{subequations}
which will be calculated numerically using
Eq. \eqref{eq:decaywidthNLOhAA}. 
The value obtained for the width depends on the renormalization scheme
used which will be discussed in the next section. 
We have explicitly checked that for all scenarios the NLO width is
UV-finite and gauge independent. 
 
 \subsection{Allowed Parameter Space}
\label{ch:evaluation}

For our numerical investigation we performed a scan in the CxSM
parameter space using \texttt{ScannerS}
\cite{Coimbra:2013qq,PhysRevD.92.025024, Muhlleitner:2020wwk} and kept
only those points that are compatible with the above described
theoretical and experimental constraints.
The scan ranges for the input parameters are summarized in
Tab. \ref{tab:scanners}. The DM mass has to be below 62.5 GeV for
$h_{125} \rightarrow AA$ to be kinematically allowed. The SM input
parameters are taken from \cite{Zyla:2020zbs}  
and their values are given in Tab. \ref{tab:inputparameters}. Note that all these parameters enter the calculation
via the EW one-loop corrections.

\begin{table}[h!]
\centering
\begin{tabular}{ccc}
\toprule
Parameter & \multicolumn{2}{c}{Range} \\
\cmidrule{2-3}
 & {Lower} & {Upper} \\
\midrule
$m_s$ & 30 GeV & 1000 GeV\\
$m_A$ & 10 GeV & 62 GeV \\
$v_S$ & 1 GeV & 1000 GeV\\
$\alpha$ & $-$1.57 &  1.57 \\
\bottomrule
\end{tabular}
\caption{The scan ranges used for the generation of parameter points
  with \texttt{ScannerS}.} 
\label{tab:scanners}
\end{table}

\begin{table}[h!]
\centering
\begin{tabular}{cc}
\toprule
SM parameter & Value \\
\midrule
$m_Z$ & 91.1876 GeV \\
$m_W$ & 80.379 GeV\\
$m_{h_{125}}$ & 125.09 GeV\\
$m_\tau$ & 1.777 GeV\\
$m_b$ & 4.7 GeV  \\
$m_t$ & 172.5 GeV\\
\bottomrule
\end{tabular}
\caption{The SM parameter values used in the numerical evaluation
  taken from \cite{Zyla:2020zbs}. 
  }
\label{tab:inputparameters}
\end{table}

We have also used the program \texttt{BSMPT} \cite{Basler:2018cwe,
  Basler:2020nrq} to check for the possibility of having a strong
first order EW phase transition (SFOEWPT). We found that in the
parameter space probed there were no points with a SFOEWPT.
Before starting the discussion of the allowed parameter space we again
remind the reader that there is a kinematical constraint that applies
to the process-dependent scheme but not to the ZEM scheme of the
counterterm $\delta v_S$. 

\begin{figure}[h!]
\centering
\includegraphics[width = 0.65 \linewidth]{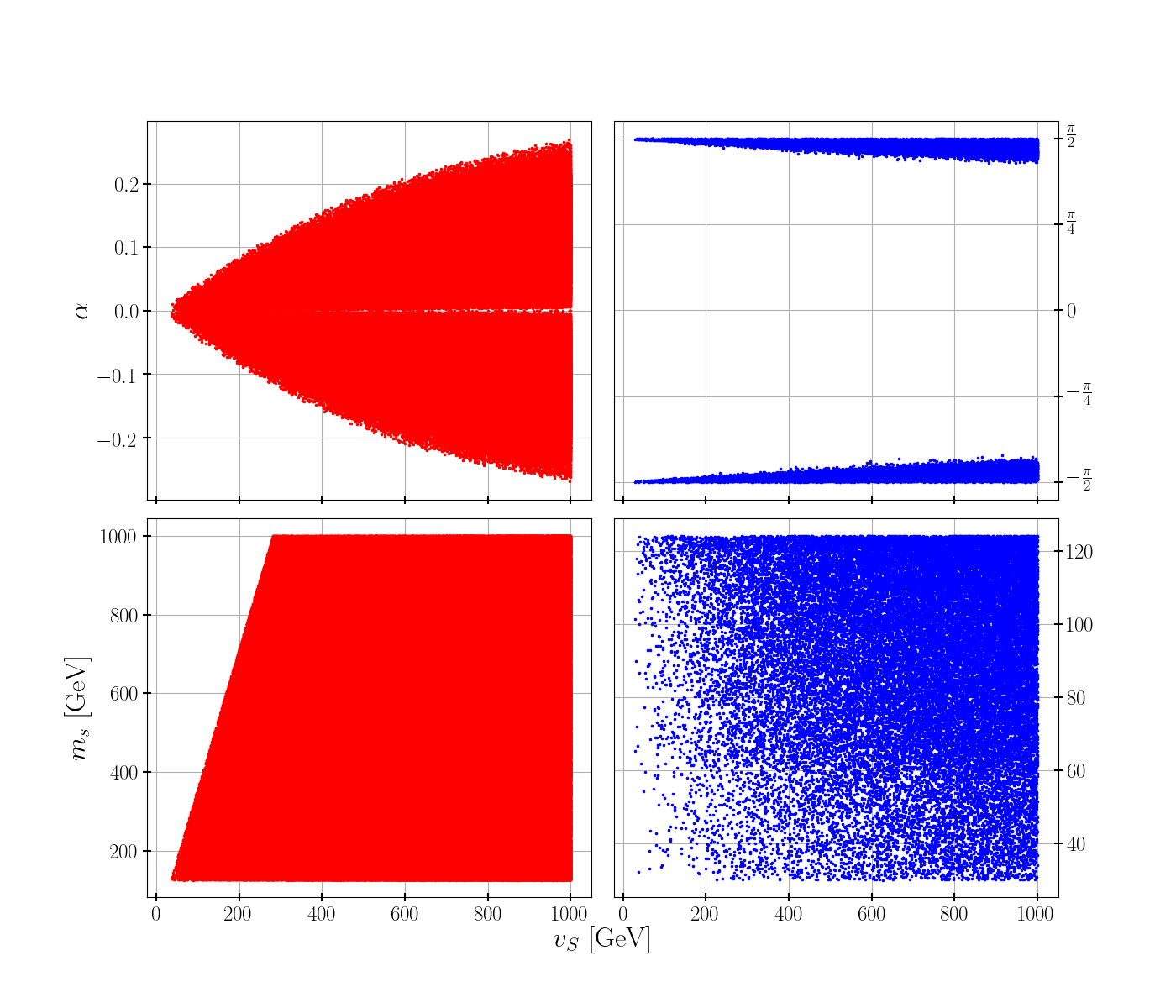}
\caption{Input parameters $\alpha$ vs. $v_S$ in the upper row and $m_s$ vs. $v_S$ in the lower row. The red plots on the
 left side are for the scenario where the 125 GeV Higgs boson is the lighter scalar particle and the blue plots on the right are for the heavier Higgs scenario.}
\label{fig:paramplots}
\end{figure}

As previously discussed two of six parameters are fixed, one by $G_F$ and the
other one is the  125 GeV Higgs boson mass. This leaves us with the 4
input parameters $m_s, m_A, \alpha, v_S$ 
where $m_s$ denotes the scalar mass of the non-125 GeV Higgs boson. In Fig. \ref{fig:paramplots} we show correlations between $\alpha$, $v_S$ and $m_s$. 
In the upper row a strong correlation can be seen between $\alpha$ and
$v_S$. This is to be expected since all SM couplings to the $h_{125}$
Higgs boson have an additional $c_\alpha$ in scenario I or $s_\alpha$
in scenario II. These couplings are very well measured and only small deviations are allowed. Thus, the additional factor has to be close to 1 and $\alpha$ has to be close to 0 or $\pm \frac{\pi}{2}$, respectively. Moreover, the parameters $\alpha$ and $v_S$ are connected through the decay width of the 125 GeV Higgs boson into DM particles. As can be seen in Eq. \eqref{eq:LOresult}, the LO decay width in scenario I is proportional to
\begin{align}
\Gamma^{\mathrm{LO}}_{h_{1} \rightarrow AA} \propto \frac{s_{\alpha}^{2}}{v_S^{2}}.
\end{align}
Thus, in order for the LO branching ratio of the 125 GeV Higgs into DM particles in the CxSM not to exceed experimental limits \cite{ATLAS:2019cid}, this ratio has to be small. Therefore, if $v_S$ is small $\alpha$ has to be small. This behavior can be seen in Fig. \ref{fig:paramplots}. In scenario  II the LO decay width is proportional to
\begin{align}
\Gamma^{\mathrm{LO}}_{h_{2} \rightarrow AA} \propto \frac{c_{\alpha}^{2}}{v_S^{2}}.
\end{align}
Therefore, if $v_S$ is small, $\alpha$ has to be close to $\pm \frac{\pi}{2}$ which can be seen in Fig. \ref{fig:paramplots} as well. One should also mention that there is a hard
bound on $\alpha$ coming from the Higgs coupling measurements. 

The plots in the lower row in Fig.~\ref{fig:paramplots} show the
relation between $v_S$ and $m_s$. The two parameters $m_s$ and $v_S$
can be related via $d_2$. Because in scenario I $m_s = m_{h_2}$ and
$\alpha$ cannot deviate much
  from zero we can write
\begin{align}
\label{eq:d2simplified}
d_2=\frac{m_{h_{125}}^2+ m_{s}^2+ \cos(2 \alpha) (m_{s}^2-m_{h_{125}}^2)}{v_{S}^2} \xrightarrow{\alpha \rightarrow 0} \frac{2 m_s^2}{v_S^2}.
\end{align}
Using again the small angle approximation in Eq. \eqref{eq:inputparameters}, $\lambda$ and $\delta_2$ can be expressed as
\begin{align}
\lambda &\xrightarrow{\alpha \rightarrow 0} \frac{2
          m_{h_1}^2}{v^2}=\frac{2 m_{h_{125}}^2}{v^2}, \\
\delta_2 &\xrightarrow{\alpha \rightarrow 0} 0.
\end{align}
With this simplified expressions the fourth constraint in Eq. \eqref{eq:unitarity} results in 
\begin{align}
\left| \frac{3}{2} \lambda + d_2 \pm \left( \frac{3}{2} \lambda - d_2 \right) \right| &\leq 16\pi 
\quad \Rightarrow d_2 \leq 8 \pi \quad  \Rightarrow  m_s \leq \sqrt{4\pi} v_S,
\end{align}
where $d_2$ was considered to be positive. This relation explains the
line in Fig.~\ref{fig:paramplots} (lower left) for scenario I, showing
$m_s$ and $v_S$ are linearly related with the correctly predicted
slope. 
The same calculation applies to scenario II. In this case, $m_s = m_{h_1}$ and the angle $\alpha$ is close to $\pm \frac{\pi}{2}$. 
The conclusion is again that $m_s$ and $v_S$ are linearly related. For example, setting $m_s$ to the highest possible value in this scenario, i.e. about 125 GeV,
 $v_S$ has to be at least 35 GeV. In this scenario only a small part
 of the parameter space is constrained but in
 Fig.~\ref{fig:paramplots} (right) we see that the far left side
 of the plot indeed contains  
 no parameter points in scenario II. 

\begin{figure}[h!]
\centering
\includegraphics[width = 0.65 \linewidth]{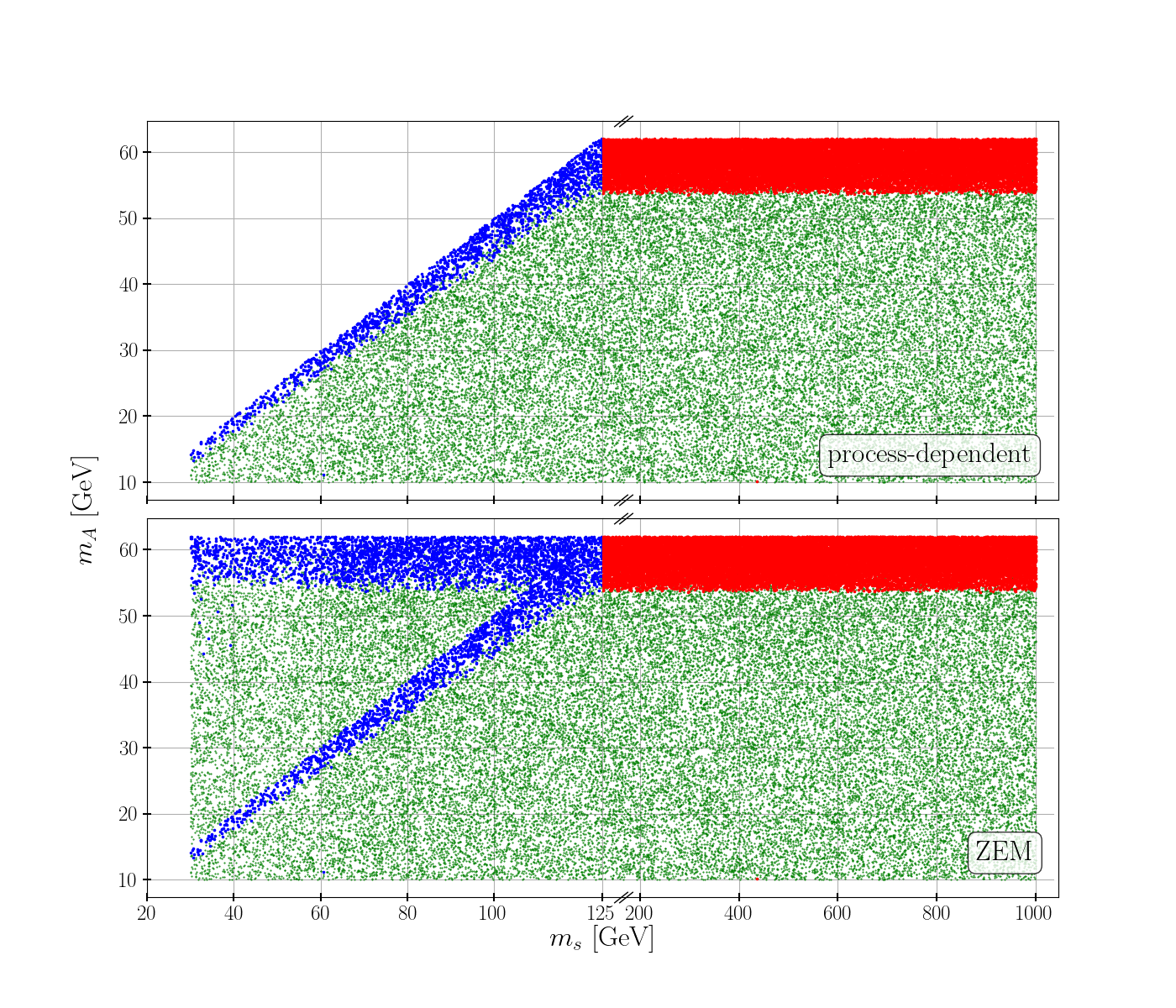}
\caption{$m_A$ vs. the non-125 GeV scalar mass $m_s$. The red points are for the scenario where the 125 GeV Higgs is the lighter scalar particle and the blue points the other scenario. 
The green points are parameter points rejected by  DM constraints.}
\label{fig:paramplotsNoMicro}
\end{figure}

Fig.~\ref{fig:paramplotsNoMicro} shows the parameter space spanned by
$m_s$ and $m_A$. The blue points (scenario II) are the ones where the
kinematical constraint (due to the process-dependent scheme) appears.  
 As expected the constraint is not there for scenario I (red
 points). In scenario I the DM mass $m_A$ prefers values close to 125/2 GeV, 
whereas in scenario II (blue points), $m_A$ has values close to half
of $m_s$ or also close to half of $m_{h_{125}}$
{in the ZEM scheme where the kinematic constraint $2m_A
 < m_s$ from the renormalization condition on $v_S$ ceases to apply. This
behavior results from DM constraints applied on the DM mass $m_A$. 
To visualize the effect of DM constraints, we show in green the points
that passed all constraints except the dark matter ones. 
The reason for these constraints is the requirement that the relic density obtained in the CxSM must not exceed the observed value of the relic density. Therefore, the thermal annihilation processes of two DM particles $A$ into one of the scalar particles 
$h_i$ must be efficient enough. This annihilation is enhanced close to the threshold, so that the DM mass $m_A$ is preferably close to half of the 125 GeV or half of $m_s$.

\begin{figure}[h!]
\centering
\includegraphics[width = 0.38 \linewidth]{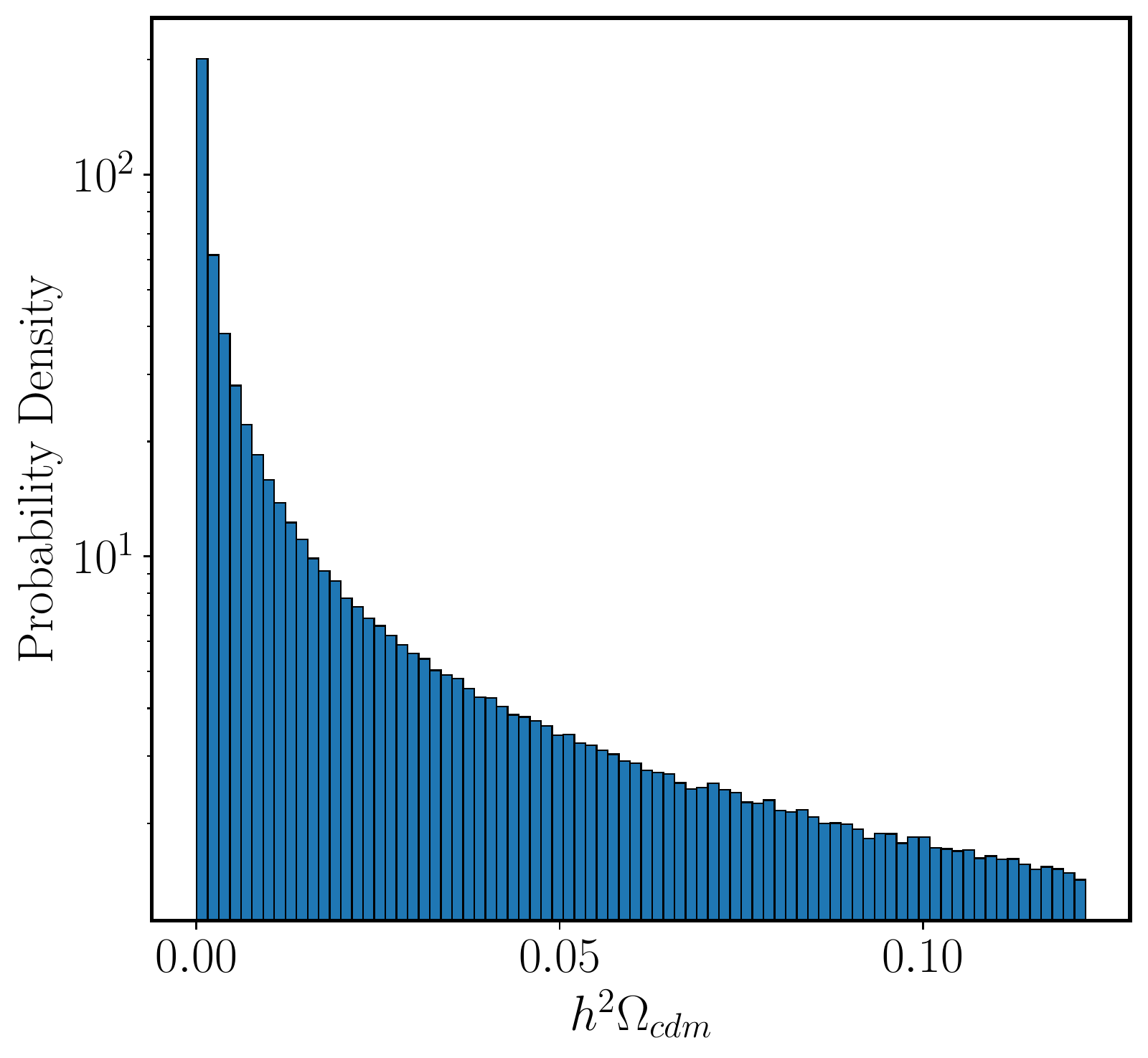}
\includegraphics[width = 0.42 \linewidth]{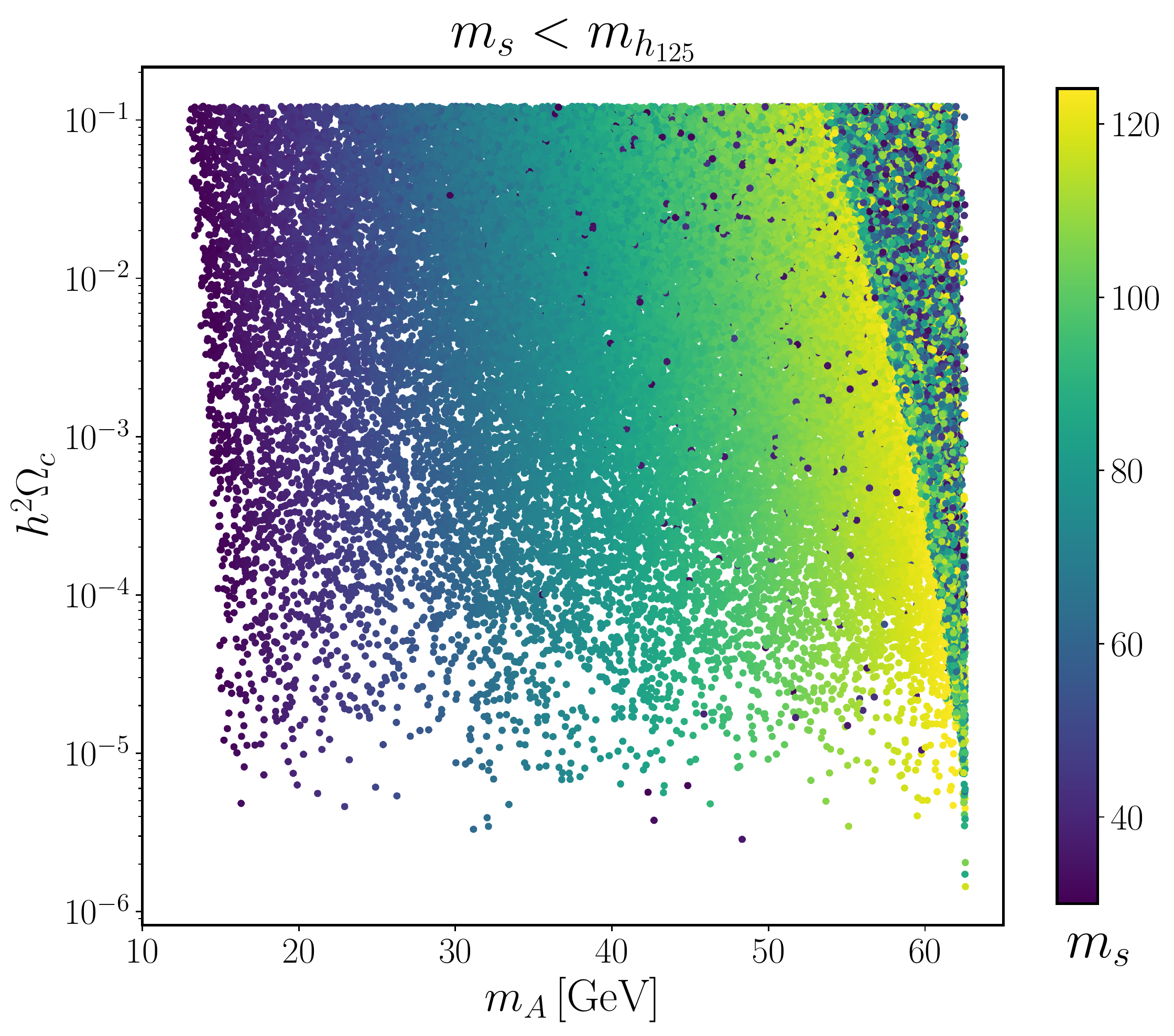}
\caption{Left: histogram showing the points frequency as a function of the relic density. Right: relic density as a function of the DM mass with $m_s$ presented by the color bar for
 the scenario where $m_{h_2} = 125$ GeV. }
\label{fig:hist}
\end{figure}

In Fig~\ref{fig:hist} (left) we present a histogram showing the
points frequency as a function of the relic density for both scenarios. This plot clearly shows us that there are points
that saturate the relic density but most of the points have a low $h^2 \Omega_{cdm}$ and would need other DM candidates. The percentage of points that
is in the range $-5\sigma < h^2 \Omega_{cdm}^{cv} \leq 2\sigma $,
where $h^2 \Omega_{cdm}^{cv}$ is the experimental central value, is around 1\% and
the preferred values for the parameters are for the two resonant regions already discussed. In the right panel we present the relic density as a function of the DM mass with $m_s$ presented by the color bar 
for the scenario where $m_{h_2} = 125$ GeV. There are points that
saturate the relic density in the entire DM mass range probed. We
clearly see that these points all have a DM mass that is half of $m_s$
or half $m_{h_2}$. There are also some outliers that saturate the relic density
in the region where $m_s$ is roughly between 30 and 50 GeV for a DM mass above 30 GeV.  
 For the other scenario, since only the case half of $125$
GeV is possible all values of $m_{h_2}$ can in principle saturate the relic
density. 

\begin{figure}[h!]
\centering
\includegraphics[width = 0.85 \linewidth]{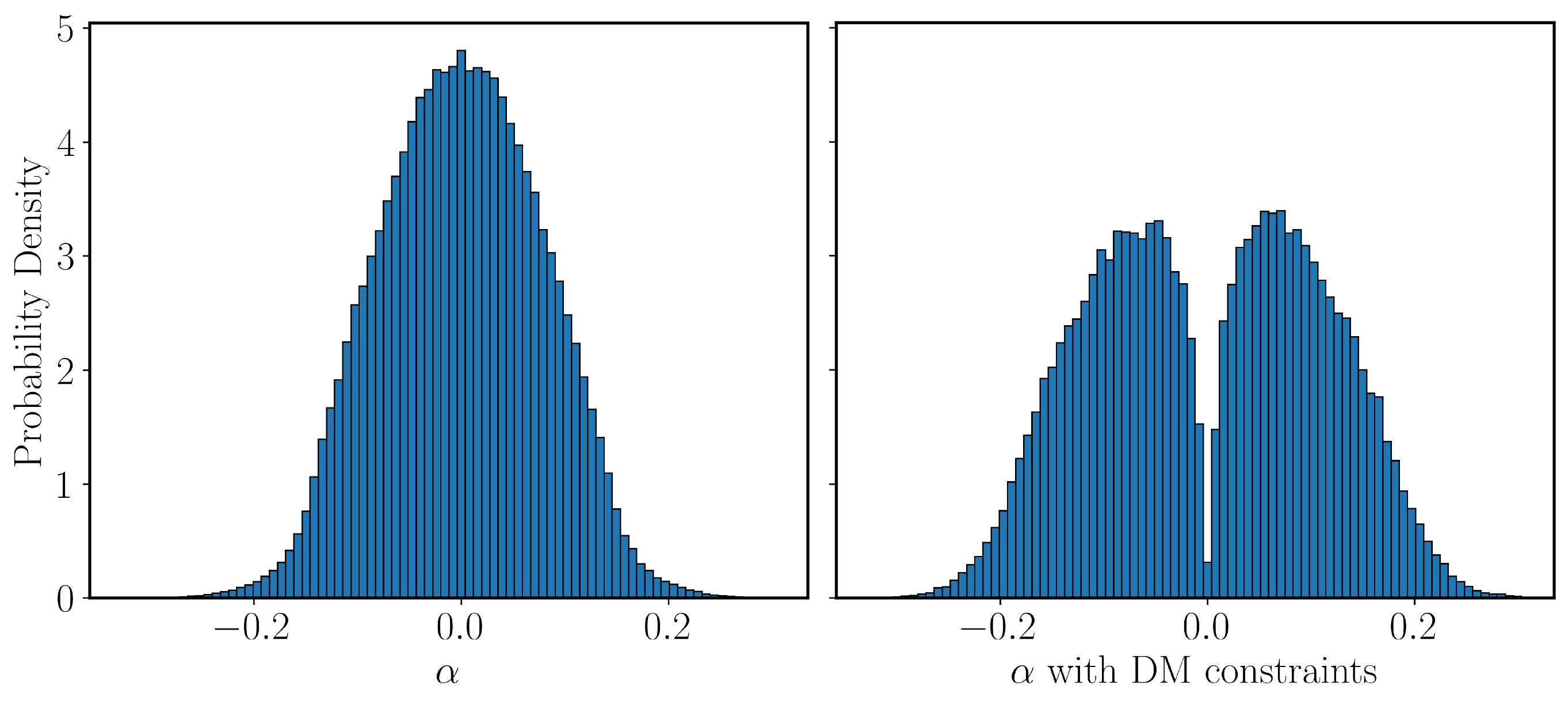}
\caption{Histogram of the frequency of the variable $\alpha$ without
  (left) and with (right) the relic density constraint for scenario
  I.}
\label{fig:histalpha}
\end{figure}

In Fig~\ref{fig:histalpha} we show a histogram of the frequency of the variable $\alpha$ without and with the relic density constraint for scenario I. Without the DM
constraints there is a bound on $\alpha$ that forces it to be close to zero. This is related to the already discussed bounds from colliders. Looking at the Boltzmann equation
\begin{equation}
\frac{dn}{dt} + 3Hn = \left< \sigma v   \right>   (n_{eq}^2 - n ^2)
\end{equation}
where $n$ is the DM number density, $H$ is the Hubble parameter,  $\left< \sigma v   \right> $   is the velocity-averaged cross section and $n_{eq}^2 $
is the density  of DM particles when in thermal equilibrium with the photon bath. The annihilation cross section $\sigma (AA \to SM SM)$, where
$SM$ are SM particles, is proportional to $\sin \alpha \cos \alpha$. Hence, if either $\sin \alpha \to 0$ or $\cos \alpha \to 0$ we get  $\left< \sigma v   \right> \to 0$
and no freeze-out will occur or the relic density will be extremely high at the end of freeze-out.  

The interesting feature is then that as we move closer to the limit where the couplings are all SM-like ($\alpha \approx 0$ is scenario I) we lose the DM candidate because of the constraints from DM. This is not surprising because
in this limit the portal coupling vanishes and freeze-out is no longer possible.

Let us now move to the last constraint coming from DM, the direct detection process. 
Since we allow DM not to saturate the relic density we need to define a DM fraction
\begin{equation}
f_{AA} = \frac{(\Omega h^2)_A}{(\Omega h^2)_{DM}^{\text{obs}}}
\end{equation}
where $(\Omega h^2)_A$ is the calculated relic density for each point in the CxSM and $(\Omega h^2)_{DM}^{\text{obs}}$ is the central
value of the experimental measurement. In the comparison with the data, we are actually comparing an effective DM annihilation cross section defined by
\begin{equation}
\sigma_{\text{eff}} = f_{AA} \sigma_{AN}
\end{equation}
where  $f_{AA}$ and $\sigma_{AN}$, the direct
  detection DM nucleon cross section, are calculated by \texttt{MicrOMEGAs}. This is because the experimental limits assume the DM candidate to make up for all of the DM abundance.

\begin{figure}[h!]
\centering
\includegraphics[width = 0.85 \linewidth]{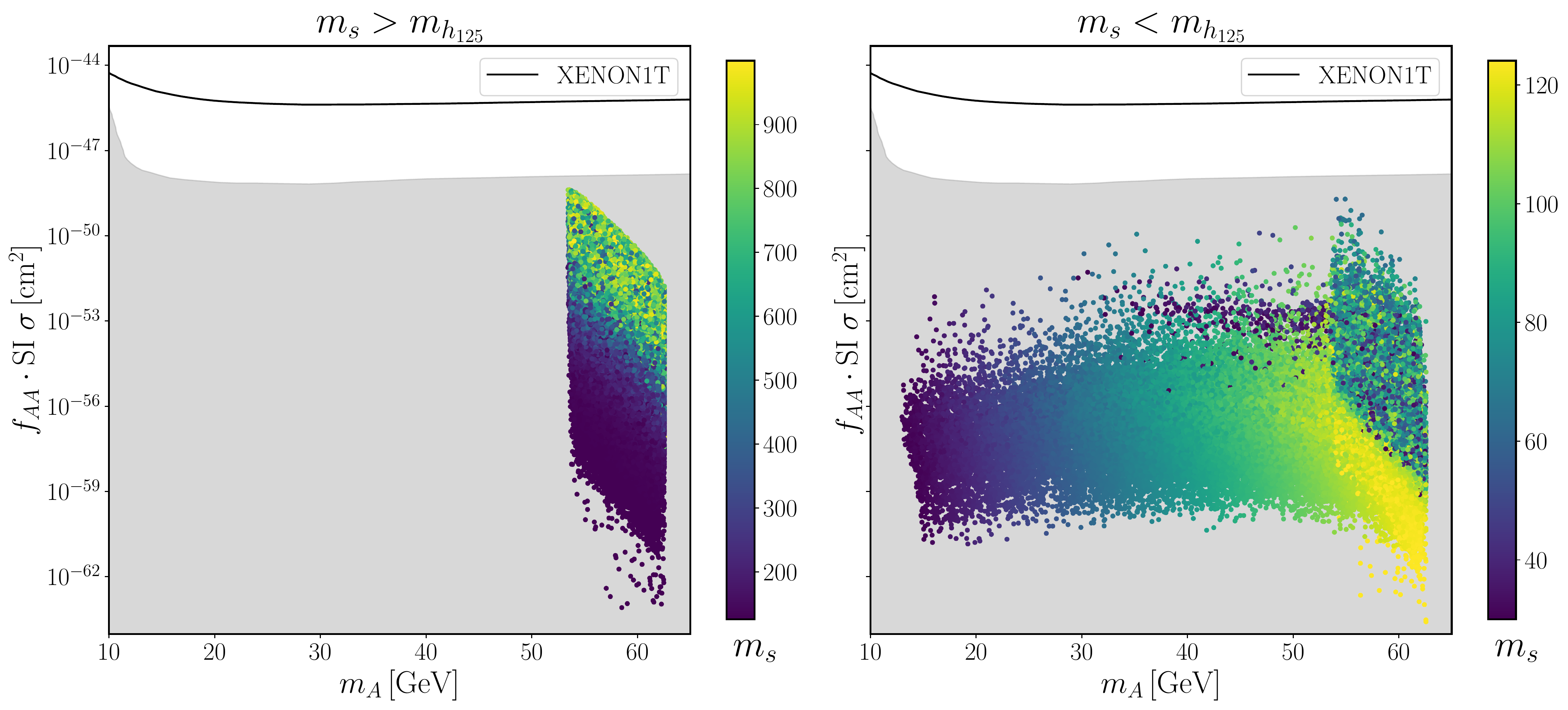}
\caption{Effective spin independent nucleon DM cross
  section as a function of 
  the DM mass for scenario I (left) and scenario II (right). Also
  shown is the XENON1T~\cite{XENON:2018voc} exclusion line (black
  line). The grey shaded region corresponds to the
    neutrino floor.}
\label{fig:direct}
\end{figure}

This constraint is particularly relevant because it directly probes
the portal coupling just like the invisible decay. Even if, as we have
already discussed, the DM nucleon cross section is only relevant at
one-loop order, it could be that the experimental bound from
XENON1T~\cite{XENON:2018voc} would provide a stronger restriction 
than the one from the invisible Higgs decay. It turns out, however, that it does not. In
Fig.~\ref{fig:direct} we present the effective
spin-independent DM nucleon cross  
section~\cite{Azevedo:2018exj, Glaus:2020ihj} as a function of the DM
mass for scenario I (left) and scenario II (right). The neutrino
floor~\cite{Billard:2013qya} is also presented as a grey shaded
region. For the range of masses considered it is below a line of about
$10^{-48}$ cm$^2$. 
We can see that the points are not only below the XENON1T line but
they are also below the neutrino floor and therefore
 have extremely
small chances of being detected directly. Therefore, in the near
future, and perhaps also in the far future, information about the dark sector
of the CxSM will come only from the LHC. This shows the importance of
 taking into account the radiative
corrections for the invisible Higgs decay.  

\subsection{Numerical Results and Analysis of the SM
    Higgs Decay into DM}
\label{sec:numresults}
In the following, we present and discuss the LO and NLO decay widths for all
allowed points in the parameter space, for the two scenarios.  
There are a total of four schemes corresponding to the combination of
the choices of the counterterms $\delta \alpha$ ($p^*$ pinched and OS
pinched) and $\delta v_S$ (process-dependent and ZEM). 
 We display results for
  the relative size of the NLO decay width with respect to the LO result, defined as
\begin{align}
\label{eq:DeltaGamma}
\Delta \Gamma \equiv \frac{\Gamma^{\mathrm{NLO}}_{h_{125} \rightarrow
  AA}}{\Gamma^{\mathrm{LO}}_{h_{125} \rightarrow
  AA}}-1 =\frac{2\mathrm{Re}\left(\mathcal{A}^{\mathrm{NLO}}_{h_{125}
  \rightarrow AA}\right)}{\mathcal{A}^{\mathrm{LO}}_{h_{125}
  \rightarrow AA}}.
\end{align}

\begin{figure}[h!]
\centering
\includegraphics[width = 0.85 \linewidth]{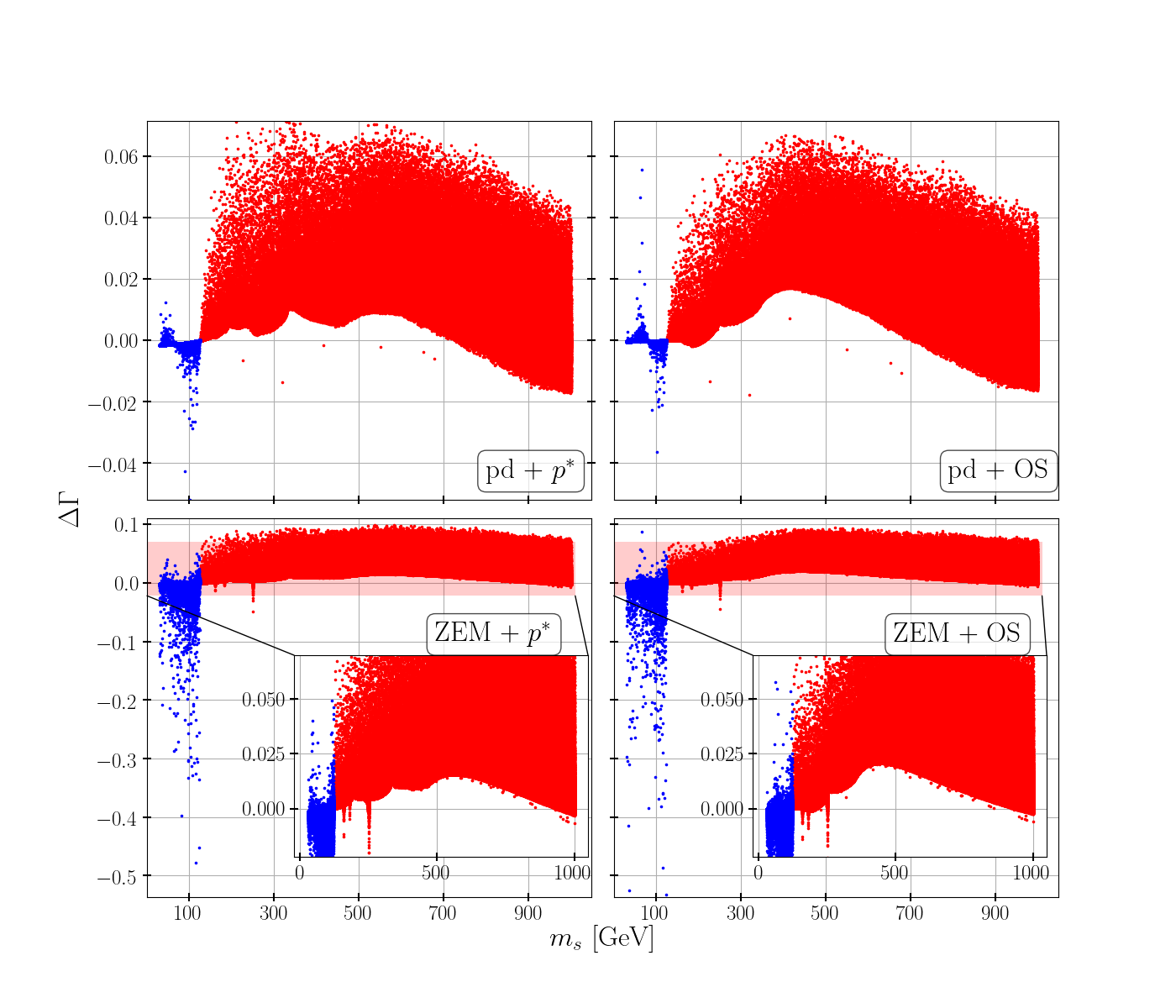}
\caption{
$\Delta \Gamma$ plotted against the scalar mass $m_s$ , where $h_{125} = h_1$ (red points) and $h_{125} = h_2$ (blue points). All different combinations of possible renormalization schemes are shown.
 Interesting sections (indicated by the red band) of the two plots in the second row are also shown in more detail.}
\label{fig:NLOLOparams}
\end{figure}

In Fig.~\ref{fig:NLOLOparams} we present $\Delta \Gamma$ as a function
of $m_s$ for the two scenarios and for the four different possible
combinations of renormalization conditions. The relative
NLO corrections in scenario II
(blue points) are quite small in the process-dependent scheme
(denoted by 'pd' in the plot), but become
comparatively large in the ZEM scheme  
with respect to scenario I (red points).
Both in scenario I and II, $\Delta \Gamma$ is barely
  affected by the choice of the renormalization scheme of
  $\alpha$. Larger differences occur when changing the renormalization
  scheme of $v_s$ from the process-dependent to the ZEM scheme but
  they still remain relatively stable in
  scenario I. Note, that the peaks in scenario I in the ZEM scheme, that
induce larger $\Delta \Gamma$ are related to 
kinematical thresholds of the $B_0$ and $C_0$ functions of the loop
integrals. They are better visualized by the zoomed inserts in
Fig.~\ref{fig:NLOLOparams}.
In scenario II, the change in $\Delta \Gamma$ when turning from the
process-dependent to the ZEM scheme has a large effect. Here,
  $\Delta \Gamma$ can go from $-50$ \%  to $10$ \%, whereas in the
  process-dependent scheme, $\Delta \Gamma$ varies between $-3$ \% and
  $3$ \%. Thus, the ZEM scheme can result in relatively large
  corrections  at NLO. These large corrections, however, only occur in
  a small number of points. These are the points
  that would be rejected by the additional kinematic constraint that in
  scenario II is effective in the process-dependent scheme. They hence
  only occur in the ZEM scheme.

One further remark is in order here. One has to be
  careful when directly comparing the results for $\Delta \Gamma$ in
  the different renormalization schemes. A consistent comparison would
require the proper conversion of the input parameters when going from
one scheme to the other. This requires the implementation of the
conversion formulae which is beyond the scope of this paper. Our goal
here primarily is to show which sizes of relative corrections at all
can be expected in the various schemes. Apart from the ZEM scheme they
are all relatively small and numerically stable in the sense defined above.

\begin{figure}[h!]
\centering
\includegraphics[width = 0.65 \linewidth]{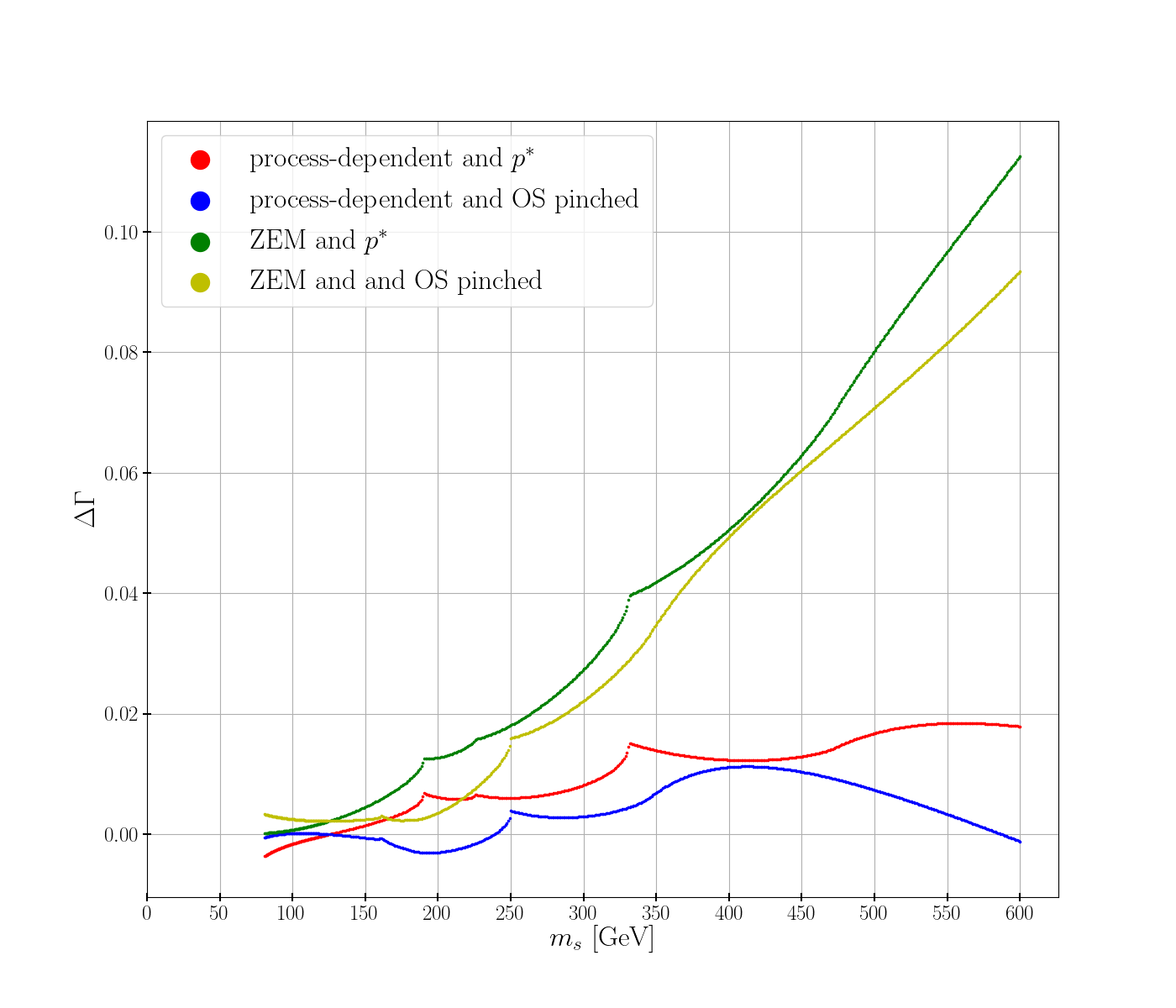}
\caption{
$\Delta \Gamma$ plotted against the scalar mass $m_s$, all other parameters have been set to fixed values, with $\alpha=0.01$, $v_S=100$ GeV and $m_A=40$ GeV. All possible combinations of renormalization schemes are shown.}
\label{fig:NLOLO}
\end{figure}

In  Fig.~\ref{fig:NLOLO} we present $\Delta \Gamma$ as a function of
$m_s$ with all other input parameters fixed. The resulting scenarios
do not necessarily fulfil all theoretical or experimental
constraints any more but are shown here for illustrative reasons. The
peaks that can be seen in the figure origin from thresholds in the
loop functions and depend on the chosen scheme as the two schemes used
for the derivation of $\delta \alpha$ are evaluated at different scales.  
For example, the peak in the OS pinched scheme seen in
Fig.~\ref{fig:NLOLO} at $m_S \equiv x_{\text{OS}}=250$~GeV appears in
the $p^*$ pinched scheme at the $m_S \equiv x_{p^*}$ value equal to 330~GeV because
\beq
x_{\mathrm{OS}}^2=\frac{m_{h_{125}}^2+x_{p^*}^2}{2},
\eeq
since in the $p^*$ pinched scheme the self-energies are evaluated at
the mean of the scalar masses.
The peaks only occur in scenario I, because most of the SM masses occurring
in the calculation (e.g. the $W$ and $Z$ boson mass) are of order of
100 GeV. 


The purpose of this analysis is to improve the precision of the calculation of the Higgs invisible decay width so that it can be used to constrain the parameters from the dark sector. 
The current observed limit on the branching ratio of the 125 GeV Higgs decay into invisible particles is given by \cite{ATLAS:2019cid}
\begin{align}
\label{eq:brlimit}
\mathrm{BR}(h_{125}\rightarrow \text{invisible}) \lesssim
  0.11^{+0.04}_{-0.03} \,,
\end{align}
at  95 \% confidence level. In order to compare results the calculated branching ratio is needed which in turn means that we need the total decay width of the 125 GeV Higgs boson in the CxSM including NLO EW corrections.
Since the corrections are not available for all decays in the model we
can only estimate the branching ratio using the total decay width of
the 125 GeV Higgs boson in the SM without EW corrections\footnote{It
  includes, however, the relevant higher-order QCD corrections that
  can be taken over from the SM to the CxSM.} which is taken from
  \cite{Djouadi_1998,Djouadi_2019} and is given by
\begin{align}
\Gamma^{\text{SM,tot}}_{h_{125}}= 0.4068 \times 10^{-2} \, \text{GeV} .
\end{align}
In order to translate this decay width into the CxSM set-up it will be multiplied by the appropriate squared angular factor $k_i^2$, where the index $i$ is chosen according to the mass scenario. Also
the NLO $h_{125} \rightarrow AA$ width is added to obtain the total decay width in the CxSM. Furthermore, in scenario II the 125 GeV Higgs boson is the heavier of the two scalar particles ($h_{125} \equiv h_2$). If  $h_1$ is light enough, the decay $h_2 \rightarrow h_1 h_1$ is also allowed and is added to the total decay width.
Thus, the LO and approximate NLO branching ratio of the decay $h_{125} \rightarrow AA$ is given by
\begin{align}
\label{eq:brapproximated}
\mathrm{BR}_{\text{CxSM}}^{\mathrm{LO/NLO}}(h_{125}\rightarrow AA) \approx \frac{\Gamma^{\mathrm{LO/NLO}}_{h_{125}\rightarrow AA}}{k_i^2 \Gamma^{\text{SM,tot}}_{h_{125}} + \Gamma^{\mathrm{LO/NLO}}_{h_{125}\rightarrow AA} + \delta \, \Gamma^{\mathrm{LO}}_{h_{125}\rightarrow h_1 h_1}} \, ,
\end{align}
where $\delta$ is defined as
\begin{align}
\delta = 
\begin{cases}
1 ,  \, m_{h_{125}} \geq 2 m_s \\
0 , \, m_{h_{125}} < 2 m_s \\
\end{cases}.
\end{align}
This expression is approximate in the sense that the NLO EW
corrections are only included in the Higgs-to-invisible decay but not
in the SM-like CxSM Higgs decays into SM particles. It is 
justified, however, if the EW corrections to these decay widths are
small enough compared to the EW corrections to the $h_{125} \rightarrow
AA$ decay\footnote{From Ref. \cite{Krause_2020}, where for the 2HDM and the
  N2HDM the EW corrections have been calculated for all the allowed
  parameter sets and in different renormalization schemes, it can be
  concluded that the EW corrections to the decay widths of the SM-like
  Higgs into SM particles amount up to a few percent only.}. Moreover,
for a better approximation the NLO corrections to the decay $h_{125}
\rightarrow h_1 h_1$ have to be included as well
unless its contribution to the total width is
  negligibly small. 
%

\begin{figure}[h!]
\centering
\includegraphics[width = 0.85 \linewidth]{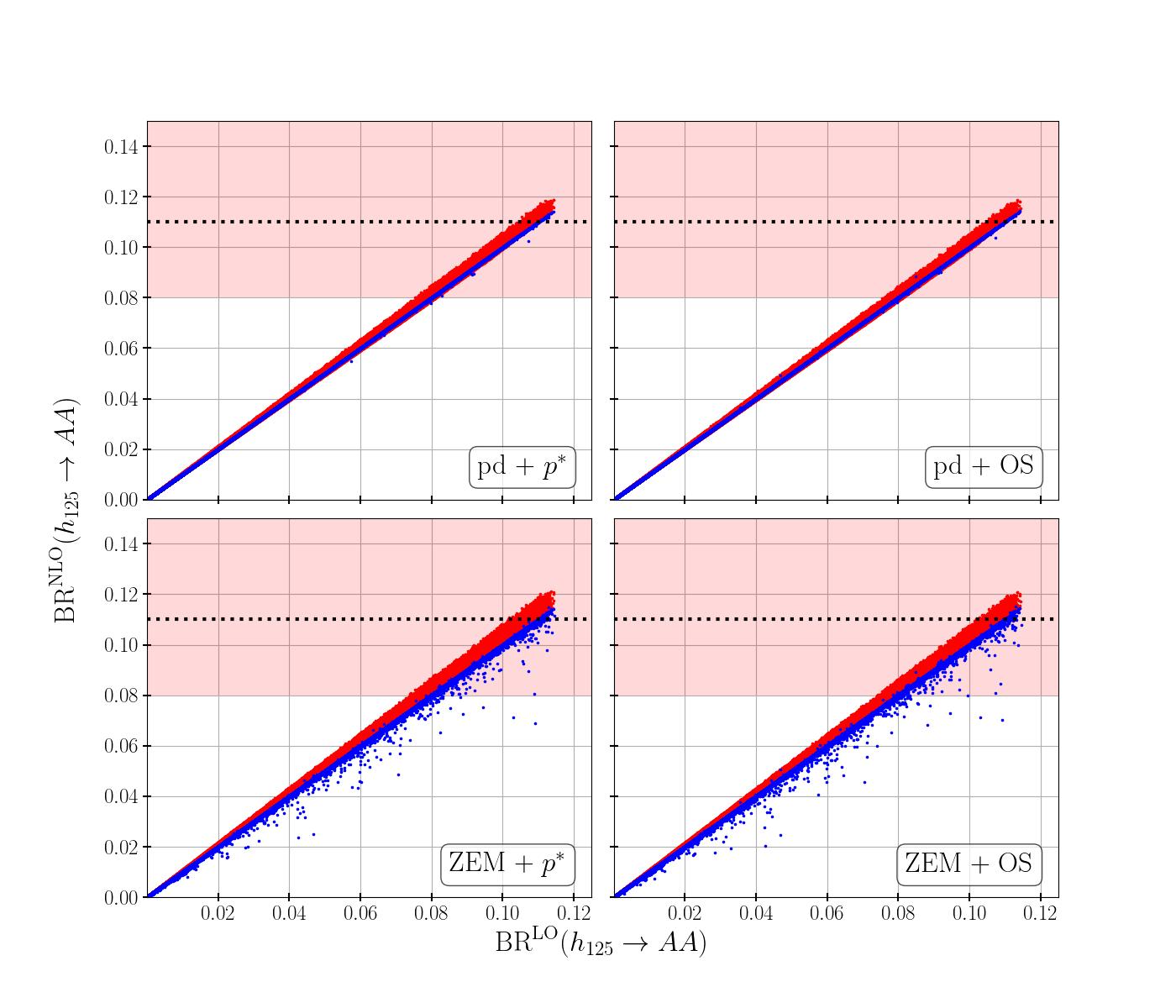}
\caption{The calculated branching ratios for the decay $h_{125} \to AA$ at NLO versus LO for all generated parameter points and all renormalization schemes. 
The experimental limit is indicated by the dashed line with the uncertainty on the limit given by the red band. Red (blue) points correspond to scenario I (II).}
\label{fig:BRBR}
\end{figure}

In Fig.~\ref{fig:BRBR} the calculated approximate NLO
  branching ratios for all 
generated parameter points are displayed versus the corresponding LO
values. The experimental limit on 
the branching ratio is shown as well. However, the limit is only
indicated for the NLO result, since the parameter points are generated
with respect to the limit at LO.  
Almost all parameter points 
have an NLO branching ratio below the experimental limit . Only about
0.2 \% of the points are above the experimental limit. The highest
obtained branching ratio is, however, around 0.121 and therefore still
lies well within the experimental uncertainty. The relative change of
the branching ratio at NLO with respect to LO has been calculated 
and increases the LO value by up
to 7-8\% at most. Thus, the NLO contributions to the
branching ratio are too 
small to further constrain the model. Moreover, it is interesting to
see that the points from scenario II result in smaller branching
ratios, especially when using the ZEM scheme. This is to be expected,
since many points in that scenario have negative relative NLO
contributions to the decay width.

\section{Conclusions} 
\label{sec:conc}

In this work we have calculated the EW NLO corrections of the Higgs decay
into two dark matter particles in the CxSM. We have used four different renormalization schemes but with all masses and fields renormalized
on-shell. Except for very particular regions of the parameter space
corresponding to thresholds in the Passarino-Veltman functions, the
corrections were shown to be quite small, on the per cent level
in all renormalization schemes. There is one
  exception, however, given by the ZEM scheme with $h_2$ being the
  SM-like Higgs. Here, points that could not be used in the
  process-dependent scheme for the renormalization of $v_S$ due to
  kinematic constraints, lead to
  relatively large corrections that amount up to a few tens of per cent.

The central value of the measured invisible Higgs branching ratio is
now at $0.11$. The inclusion 
of the EW NLO corrections  to the decay width of the process
$h_{125}\rightarrow AA$ does not lead to extra constraints on the
parameter space because the calculated approximate NLO branching
ratios for all allowed parameter  points are found to
  be within the experimental error. Calculating the EW corrections to
all decays of the SM-like CxSM Higgs boson into SM particles (and, if
kinematically allowed into a pair of lighter scalars) will
further improve the obtained result.  
But more importantly, tighter experimental constraints will be
obtained in the near future in the upcoming LHC run
\cite{Dainese:2019rgk} and even more at the high luminosity stage. 

We have also shown why it is crucial to have a precise measurement of
the invisible width - it is the only direct probe of the portal
coupling. In fact, the other possible way to probe the same coupling
would be through the DM-nucleon cross section. However, we have shown
that this cross section is not only below the present experimental
bound from XENON1T~\cite{XENON:2018voc}  
but is also below the neutrino floor which makes it virtually unusable. 
Therefore, in the near future and perhaps also in the far future,
information about the dark sector of the CxSM will come only from the
LHC.  This shows the importance of having the radiative corrections
for the invisible Higgs decay. 

\newpage
\begin{appendix}

\iflanguage{english}

\section{The Scalar Pinched Self-Energy in the CxSM}
\label{sec:pinchtechnique}
In this appendix we will present the result for the scalar pinched
self-energy in the CxSM. We define the quantity ($i,j=1,2$)
\begin{align}
O_{ij}\equiv k_i k_j,
\end{align}
to write all couplings in the CxSM between the scalars and the SM
particles $X,Y$ as
\begin{subequations}
\begin{align}
g_{XY h_i}&=g_{XY H}^{\text{SM}} k_i , \\
g_{XY h_i h_j}&=g_{XY HH}^{\text{SM}} O_{ij} ,
\end{align}
\end{subequations}
where $g_{XY H}^{\text{SM}}$ and  $g_{XY HH}^{\text{SM}}$ are the
corresponding couplings between the SM particles $X$ and $Y$ and one
or two SM Higgs bosons and $k_i$ is given in
Eq.~\eqref{eq:HiggsVbosoncoupling}. With these definitions the
self-energies $i\Sigma_{h_ih_j}^{\text{add}} $ are given by
\begin{align}
i\Sigma_{h_ih_j}^{\text{add}} (q^2) =
& \frac{-ig^2}{16 \pi^2} O_{ij} \left( q^2 - \frac{m_{h_i}^2 + m_{h_j}^2}{2} \right) B_0(q^2,m_W^2,m_W^2)  \nonumber \\
&+ \frac{-ig^2}{32 \pi^2 c_\mathrm{w}^2} O_{ij} \left( q^2 - \frac{m_{h_i}^2 + m_{h_j}^2}{2} \right) B_0(q^2,m_Z^2,m_Z^2) \nonumber \\
& + \frac{ig^2 \lambda_W}{32 \pi^2} O_{ij} \left( \left( q^2 - \frac{m_{h_i}^2+m_{h_j}^2}{2}
\right) \alpha_W - (q^4 -m_{h_i}^2m_{h_j}^2) \frac{\beta_{WW}(q^2)+\beta_{W\xi_WW}(q^2)}{2} \right) \nonumber \\
&+ \frac{ig^2 \lambda_Z}{64 \pi^2 c_\mathrm{w}^2} O_{ij} \left( \left( q^2 - \frac{m_{h_i}^2+m_{h_j}^2}{2}
\right) \alpha_Z - (q^4 -m_{h_i}^2m_{h_j}^2) \frac{\beta_{ZZ}(q^2)+\beta_{Z\xi_ZZ}(q^2)}{2} \right).
 \notag
\end{align}
Here $m_{W,Z}$ denote the masses of the $W$ and $Z$ bosons, $g=2m_W
\sqrt{\sqrt{2}G_F}$ is the $SU(2)$ gauge coupling, $c_\mathrm{w}$ the
cosine of the weak mixing angle, $\xi_{V}$ ($V=W,Z$)
are the bare gauge couplings and $\lambda_V\equiv 1- \xi_V$.
The integrals are defined as 
\begin{subequations}
\begin{align}
\frac{i}{16 \pi^2}B_0(p^2,m_1^2,m_2^2)& \equiv \int_{k} \frac{1}{(k^2-m_1^2)((k+p)^2-m_2^2)}, \\
\frac{i}{16 \pi^2}\alpha_{V}& \equiv \int_{k} \frac{1}{(k^2-m_V^2)(k^2-\xi_V m_V^2)}, \\
\frac{i}{16 \pi^2}\beta_{V_1V_2}(p^2)& \equiv \int_{k} \frac{1}{(k^2-m_{V_1}^2)(k^2-\xi_{V_1}m_{V_1}^2)((k+p)^2-m_{V_2}^2)} , \\
\frac{i}{16 \pi^2}\beta_{V_1\xi_{V_2}V_2}(p^2)& \equiv \int_{k} \frac{1}{(k^2-m_{V_1}^2)(k^2-\xi_{V_1}m_{V_1}^2)((k+p)^2-\xi_{V_2}m_{V_2}^2)}.
\end{align}
\end{subequations}

\section{Minima of the CxSM Higgs Potential}
\label{sec:minima}
To analyze all possible vacuum configurations, the scalar potential of the CxSM,
\begin{align}\label{eq:scalar_potential}
V_{\mathrm{scalar}}=\frac{m^2}{2} \Phi^{\dagger}  \Phi + \frac{\lambda}{4}\left(  \Phi^{\dagger}  \Phi \right)^2 + \frac{\delta_{2}}{2} \Phi^{\dagger}  \Phi|\mathbb{S}|^2 +\frac{b_{2}}{2}|\mathbb{S}|^2+\frac{d_2}{4}|\mathbb{S}|^4+ \left( \frac{b_1}{4}\mathbb{S}^2 +c.c.\right),
\end{align}
has to be considered with the fields defined as 
\begin{align}
\label{eq:scalar_vev_structure}
 \Phi= \begin{pmatrix}
G^+\\\frac{1}{\sqrt{2}} \left(H+iG^0 \right)
\end{pmatrix}, \:
\mathbb{S}=\frac{1}{\sqrt{2}}(S+iA).
\end{align}
Due to the $SU(2)$ invariance we can choose a configuration where only
the fields $H$, $S$ and $A$ can acquire a non-zero VEV, in the
following labeled $x_H$, $x_S$ and $x_A$.

The stationary conditions of the potential read
\begin{align}
\label{eq:VGradient}
\left. \frac{\partial V}{\partial \vec{\phi}}\right|_{\left<\phi_i \right> =x_i}=0 \; \Rightarrow \;\left\{ \begin{array}{rcc}
\frac{m^2}{2}x_H+\frac{\lambda}{4}x_{H}^3+\frac{\delta_2}{4}x_H (x_{S}^{2}+x_{A}^2)&=&0 \\
\frac{b_1+b_2}{2}x_S + \frac{d_2}{4}x_{S} (x_S^2 + x_A^2) + \frac{\delta_2}{4}x_{S}x_{H}^2&=&0\\
\frac{b_2-b_1}{2}x_A + \frac{d_2}{4}x_{A} (x_S^2 + x_A^2) + \frac{\delta_2}{4}x_{A}x_{H}^2&=&0\\
0&=&0\\
0&=&0\\
0&=&0\\
\end{array}, \right.
\end{align}
with the scalar fields collected in the vector ($G^+
  \equiv 1/\sqrt{2} (G_1 - i G_2)$)
\begin{align}
\vec{\phi}=\begin{pmatrix}
H, & S, & A, & G^0, & G_1, & G_2
\end{pmatrix}^{\mathrm{T}}.
\end{align}
The three nontrivial equations in Eq.~\eqref{eq:VGradient} can be written as
\begin{subequations}
\label{eq:minimaconditions}
\begin{align}
\label{eq:minimaH}
x_H \left( \frac{m^2}{2}+\frac{\lambda}{4}x_{H}^2+\frac{\delta_2}{4}(x_{S}^{2}+x_{A}^2) \right)=0, \\
\label{eq:minimaS}
x_S \left( \frac{b_1+b_2}{2} + \frac{d_2}{4}(x_S^2 + x_A^2) + \frac{\delta_2}{4}x_{H}^2 \right)=0,\\
\label{eq:minimaA}
x_A \left( \frac{b_2-b_1}{2} + \frac{d_2}{4}(x_S^2 + x_A^2) + \frac{\delta_2}{4}x_{H}^2 \right)=0,
\end{align}
\end{subequations}
from which we read off that for all VEVs a possible
solution is to set them to zero or solve the equations in brackets. 
Thus, eight different cases, in general, have to be
considered. Moreover, if $x_S$ and $x_A$ are 
simultaneously non-zero, the terms in brackets in 
Eqs.~\eqref{eq:minimaS} and \eqref{eq:minimaA} have to be zero. 
Since these two terms only differ in the sign in front of the parameter $b_1$,
this can only be achieved if $b_1$ is set to zero. Here, however,
$b_1$ is always chosen to be non-zero and thus these cases cannot
result in a minimum of the potential. 

Furthermore, it has to be checked whether the stationary point is
indeed a minimum of the potential, {\it i.e.}~the Hessian matrix of the
potential has to be positive definite. The general form of the Hessian
matrix reads 
\begin{align}
\label{eq:VHesse}
V_\mathrm{Hesse}=
\begin{pmatrix}
A & \frac{\delta_2 x_H x_S}{2} & \frac{\delta_2 x_H x_A}{2} & 0 & 0 & 0 \\
\frac{\delta_2 x_H x_S}{2} & B & \frac{d_2 x_S x_A}{2} & 0 & 0 & 0 \\
\frac{\delta_2 x_H x_A}{2} & \frac{d_2 x_S x_A}{2} & C & 0 & 0 & 0 \\
0 & 0 & 0 & D & 0 & 0 \\
0 & 0 & 0 & 0 & D & 0 \\
0 & 0 & 0 & 0 & 0 & D \\
\end{pmatrix},
\end{align}
where the diagonal elements are
\begin{subequations}
\begin{align}
\label{eq:diagonals}
A&=\frac{m^2}{2}+\frac{\delta_2(x_S^2+x_A^2)}{4}+ \frac{3\lambda x_H^2}{4}, \\
B&=\frac{b_1+b_2}{2} + \frac{d_2 (3x_S^2 + x_A^2)}{4} + \frac{\delta_2 x_H^2}{4}, \\
C&=\frac{-b1+b_2}{2} + \frac{d_2 (x_S^2 + 3 x_A^2)}{4} + \frac{\delta_2 x_H^2}{4}, \\
D&= \frac{m^2}{2} + \frac{\delta_2 (x_A^2+x_S^2)}{4} + \frac{\lambda x_H^2}{4}.
\end{align}
\end{subequations}

To start with the remaining cases, first the desired minimum is
considered, namely the configuration with the VEVs $x_H$ and $x_S$ to
be non-zero and $x_A$ to be zero. Since the VEVs are chosen to be
input parameters, they are in this case relabeled as $v$ and $v_S$ and
the Eqs.~\eqref{eq:minimaconditions} can be solved for other
parameters resulting in
\begin{align}
\label{eq:Eqmb1b2}
m^2=\frac{-1}{2}\left(\lambda v^2 + \delta_2 v_S^2\right), \; b_1+b_2=\frac{-1}{2} \left( d_2 v_S^2 + \delta_2 v^2 \right).
\end{align}

Next, the positive definiteness of the Hessian matrix has to be
checked. For this Eq.~\eqref{eq:Eqmb1b2} is used to simplify the
Hessian matrix in Eq.~\eqref{eq:VHesse} leading to
\begin{align}
V_\mathrm{Hesse}(x_H=v,x_S=v_S,x_A=0)=
\begin{pmatrix}
\frac{\lambda v^2}{2} & \frac{\delta_2 v v_S}{2} & 0 & 0 & 0 & 0 \\
\frac{\delta_2 v v_S}{2} & \frac{d_2 v_S^2}{2} & 0 & 0 & 0 & 0 \\
0 & 0 & -b_1 & 0 & 0 & 0 \\
0 & 0 & 0 & 0 & 0 & 0 \\
0 & 0 & 0 & 0 & 0 & 0 \\
0 & 0 & 0 & 0 & 0 & 0 \\
\end{pmatrix}.
\end{align}
The matrix is positive definite if the determinants of all minors are
positive, {\it i.e.}~the relations 
\begin{align}
\label{eq:parameterrelations}
\lambda >0 \wedge d_2>0 \wedge \lambda d_2 > \delta_2^2 \wedge b_1 <0
\end{align}
have to be satisfied. If these inequalities hold, the potential is automatically bounded from below (compare with Eq. \eqref{eq:boundedfrombelow}). Moreover, the Hessian matrix of the potential resembles the mass matrix of the scalar fields, {\it i.e.}~the eigenvalues of the matrix
are the squared masses of the corresponding particles
and thus the eigenvalues 
have to be positive, {\it i.e.}~the Hessian matrix has to be positive
definite. Furthermore, the parameter $b_1$ is just given by $- m_A^2$. 

This means that if the VEVs $v$ and $v_S$ are given as input
parameters and the VEV for the field $A$ is chosen to be zero and the
potential parameters fulfill the relations in
Eq. \eqref{eq:parameterrelations}, this configuration of VEVs is a
minimum of the potential, as desired. The remaining question now is,
whether this minimum is automatically the global minimum of the
potential. Thus, the values of the potential at all minimum
configurations have to be calculated and compared. For the desired
configuration the value of the potential at the minimum reads 
\begin{align}
V(x_H=v,x_S=v_S,x_A=0)=V(v,v_S,0)=-\frac{1}{16} (\lambda v^4 + 2
  \delta_2 v^2 v_S^2 +d_2v_S^4). 
\end{align}

Now all other VEV configurations have to be checked for their potential
values at the stationary point and whether or not they are indeed a
minimum of the potential. 
\begin{itemize}
\item case $x_H=x_S=x_A=0$:

This is the most trivial configuration, and the value of the potential at this point reads
\begin{align}
V(0,0,0)=0.
\end{align}
Thus, the difference between the values of the potential at the two configurations results in
\begin{align}
V(v,v_S,0)-V(0,0,0)= -\frac{1}{16} (\lambda v^4 + 2 \delta_2 v^2 v_S^2 +d_2v_S^4)<0.
\end{align}
The inequality is true because of the relation between $\delta_2$,
$\lambda$ and $d_2$ from
Eq.~\eqref{eq:parameterrelations}. 

\item case $x_S=x_A=0, x_H\neq 0$:

Here the nontrivial equation from Eqs.~\eqref{eq:minimaconditions} can be solved for $x_H$ and results in
\begin{align}
x_H=\sqrt{\frac{-2m^2}{\lambda}} \equiv x_1.
\end{align}
Here $m^2$ has to be negative. The value of the potential results in
\begin{align}
V(x_1,0,0)= \frac{-m^4}{4\lambda}= -\frac{(\lambda v^2 + \delta_2 v_S^2)^2}{16 \lambda},
\end{align}
where in the second step the relations Eq.~(\ref{eq:Eqmb1b2}) were used. The difference between the values of the potential of the different configurations reads
\begin{align}
V(v,v_S,0)-V(x_1,0,0)=-\frac{(d_2 \lambda - \delta_2^2)v_S^4}{16 \lambda}<0.
\end{align}
The inequality again holds because of the relations Eq.~(\ref{eq:parameterrelations}).
\item case $x_H=x_A=0,x_S\neq0$

Here the nontrivial equation from Eqs.~\eqref{eq:minimaconditions} can be solved for $x_S$ and results in
\begin{align}
x_S=\sqrt{\frac{-2(b_1+b_2)}{d_2}} \equiv x_2.
\end{align}
Here $b_1+b_2$ has to be negative. The value of the potential results in
\begin{align}
V(0,x_2,0)= -\frac{(b_1+b_2)^2}{4d_2}= -\frac{(\delta_2v^2+d_2v_S^2)^2}{16 d_2},
\end{align}
where in the second step the relations Eq.~(\ref{eq:Eqmb1b2}) were used. The difference between the values of the potential of the different configurations reads
\begin{align}
V(v,v_S,0)-V(0,x_2,0)=-\frac{(d_2 \lambda - \delta_2^2)v^4}{16 d_2}<0.
\end{align}
The inequality again holds because of the relations Eq.~(\ref{eq:parameterrelations}).

\item case $x_H=x_S=0,x_A\neq0$

Here the nontrivial equation from Eqs.~\eqref{eq:minimaconditions} can be solved for $x_A$ and results in
\begin{align}
x_A=\sqrt{\frac{-2(b_2-b_1)}{d_2}} \equiv x_3.
\end{align}
Here $b_2-b_1$ has to be negative. The value of the potential results in
\begin{align}
V(0,0,x_3)= -\frac{(b_2-b_1)^2}{4d_2}= -\frac{(4b_1 + \delta_2v^2+d_2v_S^2)^2}{16 d_2},
\end{align}
where in the second step the relations Eq.~(\ref{eq:Eqmb1b2}) were
used. Here the parameter $b_1$ does not get canceled and the
difference between the values of the potential of this configuration
with respect to the desired minimum state depends additionally on
$b_1$ and an inequality similar to the other cases cannot be shown
 as straightforwardly. It is, however,
sufficient to look at the Hessian matrix. It results in 
\begin{align}
V_\mathrm{Hesse}(0,0,x_3)=
\begin{pmatrix}
E & 0 & 0 & 0 & 0 & 0 \\
0 & b_1 & 0 & 0 & 0 & 0 \\
0 & 0 & b_1-b_2 & 0 & 0 & 0 \\
0 & 0 & 0 & E & 0 & 0 \\
0 & 0 & 0 & 0 & E & 0 \\
0 & 0 & 0 & 0 & 0 & E \\
\end{pmatrix},
\end{align}
where $E$ is a combination of potential parameters. It can be seen that $b_1$ is a negative eigenvalue of the matrix. Thus, it cannot be positive definite and this VEV configuration cannot be a minimum.
\item case $x_S=0, x_H\neq0, x_A \neq 0$

The last case is a bit more complicated, since now two VEVs are non-zero. Here it is easier to redo the same steps as in the desired minimum configuration. First, the VEVs are relabeld as $w$ and $w_A$. Next, the stationary conditions from Eqs.~\eqref{eq:minimaconditions} are solved for other parameters to obtain the relations
\begin{align}
\label{eq:paramrelationswwA}
m^2=-\frac{1}{2}(\lambda w^2 + \delta_2 w_A^2), \, b_2-b_1=-\frac{1}{2}(\delta_2 w^2 + d_2 w_A^2).
\end{align}
Similar to the last case, the value of the potential of this
configuration will again depend on $b_1$, so comparing values with the
desired minimum configuration will not lead to  a simple inequality. Thus, the Hessian matrix is again considered. With the help of Eqs.~\eqref{eq:paramrelationswwA} it can be simplified to
\begin{align}
V_{\mathrm{Hesse}}(w,0,w_A)=
\begin{pmatrix}
\frac{\lambda w^2}{2} & 0 & \frac{\delta_2 w w_A}{2} & 0 & 0 & 0 \\
0 & b_1 & 0 & 0 & 0 & 0 \\
\frac{\delta_2 w w_A}{2} & 0 & \frac{d_2 w_A^2}{2} & 0 & 0 & 0 \\
0 & 0 & 0 & 0 & 0 & 0 \\
0 & 0 & 0 & 0 & 0 & 0 \\
0 & 0 & 0 & 0 & 0 & 0 \\
\end{pmatrix}.
\end{align}
Again, $b_1$ is a negative eigenvalue of the matrix, thus it cannot be a positive definite matrix and the configuration is not a minimum.

Moreover, the similarity between the two cases with two non-zero VEVs
is interesting. If the configuration with $w$ and $w_A$ would be
chosen as the desired minimum configuration, then $b_1$ would
necessarily be positive and the minimum configuration with $v$ and
$v_S$ would no longer be a minimum. The sign in front of $b_1$ is
essentially the only difference between the fields $S$ and $A$ and
therefore also the only difference between these VEV configurations.  
\end{itemize}

To conclude, if the non-zero VEV parameters $v$ and $v_S$ are given as
input parameters and the remaining potential parameters are chosen
such that the relations Eq.~(\ref{eq:parameterrelations}) are fulfilled,
then this configuration is a minimum of the potential and it is the
global minimum (the potential is also bounded from below with the same
relations, so it really is the global minium of the potential).

\end{appendix}

\bigskip
\bigskip
\subsubsection*{Acknowledgments}
RS and JV are supported by FCT under contracts UIDB/00618/2020, UIDP/00618/2020, PTDC/FIS-PAR/31000/2017, CERN/FISPAR
/0002/2017, CERN/FIS-PAR/0014/2019. The work of FE and MM is supported by the
BMBF-Project 05H21VKCCA.

\newpage
\bibliography{HiggsInvisibleCxSM}
\bibliographystyle{JHEP}

\end{document}